\documentclass[twocolumn]{aastex62}

\providecommand{\e}[1]{\ensuremath{\times 10^{#1}}}
\newcommand{\Msun}{\ensuremath{{M_{\odot}}}}
\newcommand{\Lsun}{\ensuremath{{L_{\odot}}}}

\usepackage[T1]{fontenc}

\graphicspath{{./}{figures/}}

\received{May 8, 2020}
\revised{July 13, 2020}
\accepted{August 21, 2020}
\submitjournal{ApJ}

\shorttitle{A Census of AGN in GOODS-S/HUDF}
\shortauthors{Alberts et al.}
\begin{document}

\title{Completing the Census of AGN in GOODS-S/HUDF: New Ultra-Deep Radio Imaging and Predictions for \textit{JWST}}

\correspondingauthor{Stacey Alberts}
\email{salberts@email.arizona.edu}

\author[0000-0002-8909-8782]{Stacey Alberts}
\affiliation{Steward Observatory, 
University of Arizona, 
933 N. Cherry
Tucson, AZ 85721 USA}

\author[0000-0002-0303-499X]{Wiphu Rujopakarn}
\affiliation{Department of Physics, Faculty of Science, Chulalongkorn University, 254 Phayathai Road, Pathumwan, Bangkok 10330, Thailand}
\affiliation{National Astronomical Research Institute of Thailand (Public Organization), Don Kaeo, Mae Rim, Chiang Mai 50180, Thailand}
\affiliation{Kavli IPMU (WPI), UTIAS, The University of Tokyo, Kashiwa, Chiba 277-8583, Japan}
%\collaboration{(AAS Journals Data Scientists collaboration)}

\author[0000-0003-2303-6519]{George H. Rieke}
\affiliation{Steward Observatory, 
University of Arizona, 
933 N. Cherry
Tucson, AZ 85721 USA}

\author[0000-0002-5825-9635]{Preshanth Jagannathan}
%\altaffiliation{Creator of AASTeX v6.2}
\affiliation{National Radio Astronomy Observatory, Socorro, NM 87801, USA}
%\collaboration{(LaTeX collaboration)}

\author[0000-0003-1991-370X]{Kristina Nyland}
\affiliation{National Research Council, resident at the Naval Research Laboratory, Washington, DC 20375, USA}
%\affiliation{American Astronomical Society \\
%2000 Florida Ave., NW, Suite 300 \\
%Washington, DC 20009-1231, USA}

%\author{Others}
%\affiliation{Other Affilation}
%\affiliation{IOP Publishing, Washington, DC 20005}

%% Note that the \and command from previous versions of AASTeX is now
%% depreciated in this version as it is no longer necessary. AASTeX 
%% automatically takes care of all commas and "and"s between authors names.

%% AASTeX 6.2 has the new \collaboration and \nocollaboration commands to
%% provide the collaboration status of a group of authors. These commands 
%% can be used either before or after the list of corresponding authors. The
%% argument for \collaboration is the collaboration identifier. Authors are
%% encouraged to surround collaboration identifiers with ()s. The 
%% \nocollaboration command takes no argument and exists to indicate that
%% the nearby authors are not part of surrounding collaborations.

%% Mark off the abstract in the ``abstract'' environment. 
\begin{abstract}
A global understanding of Active Galactic Nuclei (AGN) and their host galaxies hinges on completing a census of AGN activity without selection biases down to the low-luminosity regime.  Toward that goal, we identify AGN within faint radio populations at cosmic noon selected from new ultra-deep, high~resolution imaging from the Karl~G.~Jansky~Very~Large~Array at 6 and 3~GHz.  These radio data are spatially coincident with the ultra-deep legacy surveys in the GOODS-S/HUDF region, particularly the unparalleled \textit{Chandra} 7~Ms X-ray imaging.  Combined, these datasets provide a unique basis for a thorough census of AGN, allowing simultaneous identification via (1) high X-ray luminosity; (2) hard X-ray spectra; (3) excess X-ray relative to 6~GHz; (4) mid-IR colors; (5) SED fitting; (6) radio~excess via the radio-infrared relation; (7) flat radio spectra via multi-band radio; and (8) optical spectroscopy.  We uncover AGN in fully half our faint radio sample, indicating a source density of one AGN~arcmin$^{-2}$, with a similar number of radio-undetected AGN identified via X-ray over the same area.  Our radio-detected AGN are majority radio-quiet, with radio emission consistent with being powered predominantly  by star~formation.  Nevertheless, we find AGN radio signatures in our sample: $\sim12\%$ with radio~excess indicating radio-loud activity and $\sim16\%$ of radio-quiet AGN candidates with flat or inverted radio spectra.  The latter is a lower limit, pending our upcoming deeper 3~GHz survey.  Finally, despite these extensive datasets, this work is likely still missing heavily obscured AGN. We discuss in detail this elusive population and the prospects for completing our AGN census with \textit{JWST}/MIRI.

\end{abstract}

%% Keywords should appear after the \end{abstract} command. 
%% See the online documentation for the full list of available subject
%% keywords and the rules for their use.

%% From the front matter, we move on to the body of the paper.
%% Sections are demarcated by \section and \subsection, respectively.
%% Observe the use of the LaTeX \label
%% command after the \subsection to give a symbolic KEY to the
%% subsection for cross-referencing in a \ref command.
%% You can use LaTeX's \ref and \label commands to keep track of
%% cross-references to sections, equations, tables, and figures.
%% That way, if you change the order of any elements, LaTeX will
%% automatically renumber them.
%%
%% We recommend that authors also use the natbib \citep
%% and \citet commands to identify citations.  The citations are
%% tied to the reference list via symbolic KEYs. The KEY corresponds
%% to the KEY in the \bibitem in the reference list below. 

\section{Introduction} \label{sec:intro}

Compelled by evidence such as the link between black hole (BH) mass and bulge properties in local galaxies \citep[i.e.][]{mag98}, efforts to determine the extent of the co-evolution between galaxies and BH growth via Active Galactic Nuclei (AGN) remain at the forefront of galaxy evolution studies. This link appears to be fundamental, since it holds across multiple epochs, with the volume-averaged cosmic star formation and BH accretion histories both peaking at cosmic noon ($z\sim1-3$), with a strong decline toward the present day \citep{mad14}. 

Definitive evidence for a causal connection has been elusive, however. For example, \citet{diamond2012} studied the Shapley-Ames Seyfert sample (low-luminosity, optically selected AGN) and found a correlation between the black hole accretion rate (BHAR) and circumnuclear star formation (SF), but little such correlation for the total SF in the host galaxy. \citet{chen13} evaluated a more luminous mid-IR color and X-ray selected sample, which showed a strong correlation between the BHAR and total host galaxy SF, a result that was confirmed by others  \citep[i.e.][]{del15, lanz17}. On the other hand, \citet{xu2015} demonstrated, via a mid-IR selected sample with spectroscopic follow-up, that an apparent correlation between SF and peak BHAR can arise from the dependence of each parameter on stellar mass $-$ the SF through the Main Sequence, and the BHAR through the Magorrian relation $-$ but without necessarily a more direct causal connection. This possibility had previously been suggested by \citet{rafferty11} and is supported by, e.g., \citet{yang17}. Although both processes must be linked by a dependence on a common gas supply \citep[i.e.][]{dim05, smo08, hir14, vit14, dai18}, these studies demonstrate that the details of their connection are not well understood observationally.  

An attractive concept is that an evolutionary sequence links star forming galaxies (SFGs) and rapidly accreting black holes (i.e. AGN), progressing from Ultraluminous Infrared Galaxies (ULIRGs) to obscured AGN to unobscured AGN to quiescent galaxies \citep{hop06}. Direct and definitive  evidence has been difficult to find, however, and the association of luminous AGN with main sequence SFGs may even contradict the simplest form of this sequence \citep{xu2015}. Unfortunately,  AGN-galaxy co-evolution studies meant to quantify the SF/AGN relationship are fundamentally impeded by the difficulties in obtaining a complete census of AGN activity.  It is now well-established that all AGN selection techniques suffer from incompleteness and bias, typically driven by obscuration and viewing angle \citep[i.e.][]{jun11, cap14, del17}.  

To combat this issue, it is necessary to take advantage of the full (X-ray to radio) galaxy spectrum for AGN identification. 
Furthermore, survey depth strongly effects the outcome; deeper data, achieving better completeness at and below the boundaries between SF and AGN dominated emission at a given wavelength, generally reveal previously unknown AGN.
In this work, we extend AGN identification by complementing the existing ultra-deep legacy datasets in the GOODS-S/{\it Hubble} Ultra Deep Field (HUDF) region with new ultra-deep radio data. 

The radio spectrum provides a unique, yet complicated, tool in AGN identification and AGN-galaxy co-evolution studies.  At the bright end ($\gtrsim$ 1 mJy at z $\gtrsim$ 1), low frequency ($<$ 30 GHz) radio sources are universally understood to be dominated by bright, non-thermal synchrotron emission generated via AGN \citep[i.e.][]{mig08}.  Typically, this synchrotron spectrum is optically thin and steep, associated with radiatively inefficient accretion activity \citep{mei02, jes05, fan11, pad16, man17}.  However, radiatively efficient thin disk accretion onto particularly massive black holes may produce a flat radio spectrum, such as in the case of blazars \citep{ghi13, pad17}.  Collectively, these bright radio sources are termed ``radio-loud" (RL) or jetted, identified via radio morphological features and/or elevated radio emission relative to wavelength regimes that probe stellar emission \citep[see][and references therein]{hec14, kel16, pad16}.

The radio picture is even more complex at the faint end.  A flattening of the radio number counts below 1 mJy \citep[e.g.][]{con84, win85, hop98, ric00, sey04, sim06, kel08, owe08, con12} points to distinct radio population(s), a mix of ``radio-quiet'' (RQ) AGN\footnote{Various other terms have been introduced to be more descriptive of possible mechanisms operating in these sources, e.g., ``non-jetted'' \citep{pad16} or ``core-dominated'' \citep{whit17}.} and SFGs.  In this regime, AGN are often by necessity identified at non-radio wavelengths and the mechanism(s) responsible for the bulk of the radio emission is unclear. There are multiple lines of evidence pointing to the radio spectra of most RQ AGN being dominated by star formation, including radio-to-infrared ratios that are similar to that of SFGs \citep{bon13, bon15, pad15} as well as similarities in host galaxy colors, optical morphologies, and stellar mass between the SFG and RQ AGN populations \citep[i.e.][]{kim11, con13, kel16}. On the other hand, studies of luminous RQ AGN (i.e. quasars) find elevated radio emission relative to expectations from current SFRs \citep{zak16, white2017}, suggesting  that a mix of star formation and nuclear activity contributes to the radio outputs of these sources.  High-resolution radio observations of RQ quasars and AGN find that the radio cores of such AGN contribute significantly to but do not dominate the total radio outputs \citep{jac15, mai16, her16}.  Disentangling the sources of radio emission in the faint radio population is a necessary step toward utilizing the radio properly as an extinction-free SFR indicator and measure of AGN activity.

This paper explores AGN at the faint end of radio populations at cosmic noon ($z\sim1-3$), utilizing VLA imaging at 6 and 3 GHz at unprecedented depths (0.32 and 0.75  $\mu$Jy beam$^{-1}$, respectively, at the pointing center).  Combined with the ultra-deep legacy datasets in GOOD-S/HUDF, we identify and characterize AGN within a radio-selected sample, using multiple techniques:  these include the  radio-infrared relation, the presence of a flat slope between 3 and 6 GHz, and a high ratio of X-ray to 6 GHz luminosity. These methods are combined with purely X-ray identifications, the presence of mid-IR excesses, and optical spectroscopy, where available, to derive the most complete sample of AGN possible with the current data.  We classify these AGN into radio-loud and radio-quiet, and show that the radio emission of the RQ AGN in this faint sample is usually dominated by star formation.  Combining this radio-selected sample of AGN with X-ray AGN in radio-undetected galaxies over the same field allows us to put a lower limit on the number density of AGN.  Finally, we broaden the discussion to what type of AGN may still be missing from our census and analyze the prospects for future detections with the {\it James Webb Space Telescope}, which will open up new avenues into identifying the most obscured AGN.

The paper is structured as follows: Section~\ref{sec:data} presents the sample selection based on new VLA 6 GHz imaging, cross-matched with new 3 GHz imaging, existing ancillary data, and measured galaxy properties.  Section~\ref{sec:radiosources} follows with a brief run down of the demographics of our radio-selected sample.  In Section~\ref{sec:agn}, we identify AGN within our radio sample via multiple methods: X-ray based selections, mid-infrared (MIR) excess via colors and spectral energy distribution (SED) fitting and decomposition, and radio signatures.  We additionally identify AGN in radio-undetected galaxies over the relevant area using X-ray properties.  A discussion of our results is provided in Section~\ref{sec:disc}, including the nature of radio emission in galaxies containing AGN, the source density of AGN at cosmic noon, and prospects for identifying and characterizing the most heavily obscured AGN, a population still largely missing from current surveys.  Section~\ref{sec:con} contains our conclusions.  Throughout this paper, we adopt the convention $S_{\nu}\propto \nu^{\alpha}$ for the radio spectral slope, where $S_{\nu}$ is the radio flux density and $\alpha$ is the spectral slope.  We assume a \citet{cha03} IMF and ($\Omega_{M}, \Omega_{\Lambda}, h$) = (0.27, 0.73, 0.71) \citep{spe03}.

\section{Data} \label{sec:data}

\subsection{Radio Observations} \label{sec:radiodata}

Our main datasets are radio imaging obtained at 6 GHz (5 cm) and 3 GHz (10 cm) from the Karl G. Jansky Very Large Array (VLA).  For the former, 177 hours of C-band imaging ($4-8$ GHz) centered on the HUDF ($\alpha, \delta=\,3$:32:38.6, $-$27:46:59.89) was obtained from March 2014$-$September 2015 in A, B, and C configurations. Data reduction was performed as described in \citet{ruj16} using the {\tt Common Astronomy Software Applications (CASA)} package \citep{mcm07}.  The final map has a 0.31\arcsec x0.61\arcsec synthesized beam and rms noise of 0.32$\,\mu$Jy beam$^{-1}$ at the pointing center.  
%For further details, see \citet{ruj16}.

3 GHz imaging was obtained in a single pointing with the same pointing center as above.   This pointing was observed for 90 hours in A, BnA, and B configurations during January$-$June 2018 using the S-band receiver covering $2-4$ GHz.  The observations comprised 36 dynamically-scheduled sessions of $1.5-4$\,hours.  Each session observed 3C48 for flux and bandpass calibrations; J0402-3147 was observed for phase calibration every 25 minutes.  The data reduction and extracted source catalog for the 3 GHz survey will be presented in Rujoparkarn, in prep; we summarize the data reduction here. Calibration and imaging were done with {\tt CASA} using the following steps: 1) calibration and flagging of data using the VLA Data Reduction Pipeline (Chandler et al., in prep); 2) removal of any portions of the data corrupted by strong radio frequency interference; and 3) imaging with the task {\tt TCLEAN}. The imaging parameters were the following: MT-MFS deconvolver with nterms of 2 and Briggs weighting with robust parameter of 0.5, but with a pixel size of $0.15\arcsec$. We imaged the 3 GHz data well beyond the primary beam radius of $7.2'$ (employing the $w$-projection with {\tt wprojplanes} of 128) to mitigate the imaging artifacts caused by the sidelobes from bright sources far from the pointing center. A wideband primary beam correction was applied to the 3 (and 6 GHz) images using the {\tt CASA} task {\tt WIDEBANDPBCOR}.  The final 3 GHz image has a $0.6\arcsec \times 1.2\arcsec$ synthesized beam and rms noise at the pointing center of $0.75\,\mu$Jy beam$^{-1}$.

For this work, source candidates were extracted from the 6 GHz map using the {\tt Python Blob Detection and Source Measurement (PyBDSM)} software package \citep[v1.8.6; ][]{moh15} down to 4$\sigma$ on both the native resolution map and one with 300k$\lambda$ tapering applied.
%to aid in the extraction of extended sources. 
The tapered map has a synthesized beamsize of 0.7$\arcsec$  with a point source sensitivity reduced by $<$ 6\% compared to the native resolution map. Tapering is utilized to recover extended emission and mitigate the bias against less compact sources, generally SFGs \citep{gui17}. We expect the typical size of the radio-emitting region of a SFG at $z\sim2$ to be $\sim0.5\arcsec$ in diameter \citep{ruj16}. Source fluxes are measured from the native or tapered map according to which maximizes the signal-to-noise ratio (SNR), defined as the ratio of the peak flux to the local rms.  All 6 GHz sources were identified within the half-power radius ($r=220^{\prime\prime}$) of the primary beam.   Source extraction was performed on the 3 GHz image using the same technique at the native resolution.  The 3 GHz catalog will be presented in Rujopakarn, in prep.  

\subsection{Optical-Far-Infrared Counterparts and Final Radio Sample} \label{sec:otherdata}

6 GHz sources are confirmed via optical/near-infrared (NIR) counterparts, obtained by matching the 6 GHz catalog to the GOODS-South 3D-HST (v4.1) photometric catalog \citep{bra12, ske14}. This catalog provides coverage over 0.3-8$\,\mu$m (observed) compiled from 20 photometric bands.  To be considered in this study, we require each 6 GHz radio source to have an optical/NIR counterpart in 3D-HST and be at $z>0.75$. We use a search radius of 0.5\arcsec\footnote{Counterpart matching to 3D-HST catalogs is done after a systematic WCS offset correction has been applied.  For more details, see \citet{ruj16}.}, though we note that all counterparts are found within 0.2\arcsec.  Our search radius is chosen to encompass the offsets found between the optical and radio peaks \citep{ruj16} and is similar to the choice made in similar surveys \citep[e.g.,][]{gui17}, while the redshift cut off is to focus our study on cosmic noon. Given the source density of 3D-HST sources and our $>4\sigma$ detection criterion, the predicted number of false detections is $<$ 0.5, i.e., negligible\footnote{ This paper is focused on sources  showing evidence of having active nuclei, further reducing any chance of false identifications (e.g., most of them have X-ray counterparts).}. After one source is manually rejected due to interference by a bright radio jet from a nearby galaxy, our final 6 GHz sample includes 100 radio sources at $z>0.75$\footnote{140 6 GHz sources total are matched to 3D-HST counterparts.  Only the 100 at $z>0.75$ are considered in this study.} over an area of 42 arcmin$^{-2}$.   We then match the 3 GHz catalog (Rujopakarn, in prep) to this sample, finding 74 counterparts within a radius of 0.5$^{\prime\prime}$.

Mid- and far-infrared (FIR) counterparts are obtained from the {\tt Rainbow} Cosmological Database through the $Rainbow$ $Navigator$ \citep{per08, bar11a, bar11b}.  In the GOODS-S/CANDELS region \citep{gro11}, {\it Spitzer}/MIPS 24$\,\mu$m from the GOODS Legacy Program \citep{mag09} reaches a nominal $5\sigma$ depth of 20$\,\mu$Jy, while the GOODS-{\it Herschel} Survey provides imaging at 70-500$\,\mu$m with a $3\sigma$ depth of 2.4 mJy at 160$\,\mu$m \citep{elb11}. Photometry in the {\tt Rainbow} database at mid-to-far-IR wavelengths was performed using {\it Spitzer}/IRAC priors as described in \citet{per10, bar11a, bar11b, raw16, rod19}. This method of photometric extraction in principle deblends the IR photometry at the resolution of IRAC (2$^{\prime\prime}$); however, to be conservative, we perform a visual inspection and flag any IR fluxes as `blended' if there is evidence of multiple radio sources contributing to the 24$\,\mu$m beam ($6^{\prime\prime}$).   Of the 100 6 GHz sources, 87 are detected at 24$\,\mu$m and 44 at 160$\,\mu$m. 11/87 MIPS counterparts are flagged as blended. 

\subsection{X-ray Imaging} \label{xraydata}

Our radio surveys are coincident with uniquely deep X-ray imaging from the {\it Chandra X-ray Observatory}.  We utilize the 7\,Ms X-ray imaging and catalog described in \citet{luo17}, matched to our radio sample using a 2$^{\prime\prime}$ search radius.  The full band (0.5-7 keV), soft band (0.5-2 keV), and hard band (2-7 keV) catalogs have limiting fluxes of $1.9\e{-17}$, $6.4\e{-18}$, and $2.7\e{-17}$ erg cm$^{-2}$ s$^{-1}$, respectively.  46/100 6 GHz sources are detected in at least one X-ray band and we adopt the X-ray luminosities, hardness ratios, and source classifications as described in \citet{luo17}.

\subsection{Measured Galaxy Properties}

Redshifts and stellar mass measurements are adopted from the recent grism spectroscopy release for GOODS-S from 3D-HST \citep[v4.1.5;][]{mom15}.  Redshifts are taken from the ``zbest" catalogs, with the following priority: spectroscopic redshifts, robust grism redshifts, and, if the former are not available, photometric redshifts determined using the full UV-NIR photometric SED and the EAZY photometric redshift code \citep{bra08}.    Stellar mass measurements were derived by the 3D-HST team using the best available redshift and the full photometric dataset using FAST \citep{kri09}.  Visual morphologies are adopted from \citet{kar15}.

\begin{figure}[ht!]
\includegraphics[scale=0.62, trim=2mm 0 4mm 0, clip]{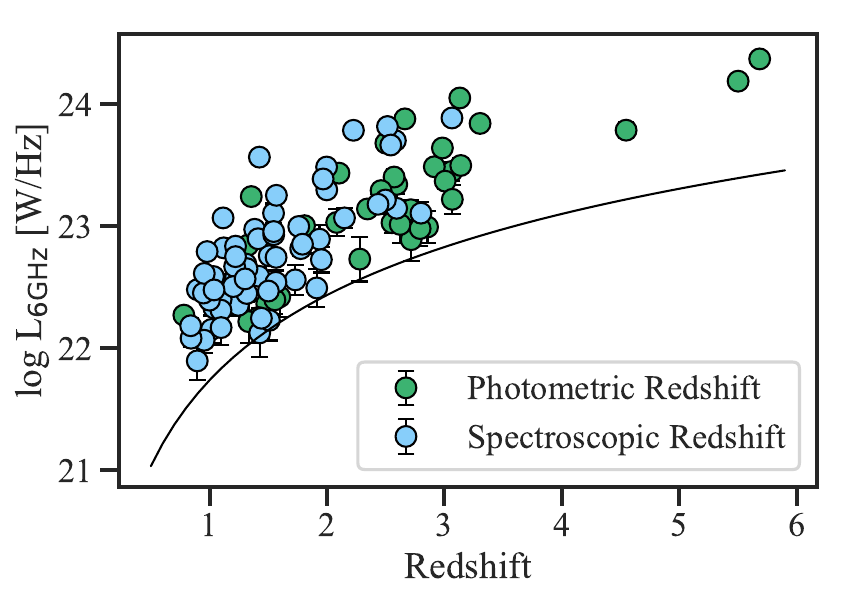}
\includegraphics[scale=0.62, trim=2mm 0 2mm 0, clip]{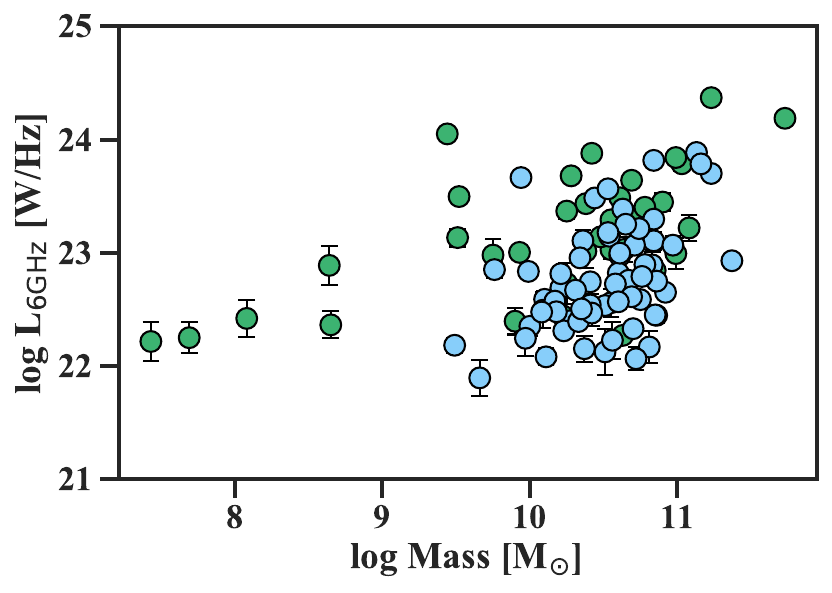}
\caption{The rest-frame 6 GHz luminosities of our sample, assuming $\alpha=-0.7$, as a function of redshift (top) and stellar mass (bottom).  Spectroscopic redshifts are indicated in blue, photometric redshifts in green. The solid line indicates the $4\sigma$ detection limit. The five cases with low (log M$_{\star}/\Msun<9$) stellar masses  have no effect on our analysis.}
\label{fig:radio}
\end{figure}

\section{Radio Source Demographics} \label{sec:radiosources}

The 6 GHz luminosities can be seen as a function of redshift and stellar mass for our 100 radio sources in Figure~\ref{fig:radio}.  The radio luminosity was calculated assuming a typical radio spectral slope, $S_{\nu}\propto\nu^{\alpha}$ with $\alpha=-0.7$, appropriate for non-thermal synchrotron emission from star forming regions \citep{con92}.  We utilize our 6 and 3 GHz datapoints to measure the true $\alpha$ for our sources in Section~\ref{sec:index}.

Using this convention, our radio sample spans a range of  $L_{\rm 6 GHz}$ $\sim 10^{22} - 10^{24}$ W Hz$^{-1}$, consistent with the radio emission expected from  (Ultra) Luminous Infrared Galaxies ((U)LIRGs) in the absence of an AGN. Most of our sample fall between $z=0.75$ and $z\sim3$, the former by design to focus on cosmic noon and the latter due to the positive k-correction in the radio.  The sample has typical stellar masses of $\sim$ $10^{9.5} - 10^{11}\,\Msun$.

\section{AGN Identification} \label{sec:agn}

One of our objectives is to gain as complete a census of AGN within our radio sample as possible. To do so, we identify AGN in the following ways: via X-ray properties 
 \citep[i.e.][]{leh10, xue11}, MIR excess via colors \citep[i.e.][]{lac04, ste05, alo06} and SED decomposition \citep{assef08, assef10, del17}, through radio properties such as radio loudness, a flat radio spectrum, and/or a radio morphology suggestive of jets \citep[see][for a review]{pad17}, and of course through optical spectra.  These multiple approaches are necessary to mitigate the selection biases inherent in each technique, typically stemming from obscuration or the AGN/host galaxy configuration \citep[i.e.][]{jun11, cap14, del17}.  In the following sections, we apply these criteria to our sample of radio sources to extract a comprehensive sample of AGN.  An executive summary of our sample and AGN identifications can be found in Table~\ref{tbl:summary} and a list of our radio sources satisfying one or more of these AGN criteria is provided in Table~\ref{tbl:simple}.

\begin{table}
\centering
\caption{ Executive summary of our sample and the AGN identified by method.  A full list is provided in Table~\ref{tbl:simple}.  All 6 GHz sources in the parent sample are required to have a 3D-HST counterpart and be at $z>0.75$.}
\begin{tabular}{lcc}
\hline
\hline
\multicolumn{2}{l}{\bf{Sample}} \\
\hline
\hline
 & Number & Section(s) \\
 \hline
6 GHz (parent sample) & 100 & \ref{sec:radiodata}, \ref{sec:otherdata} \\
\hspace{5mm} 3 GHz counterparts  & 74 & \ref{sec:otherdata} \\
\hspace{5mm} MIPS 24$\,\mu$m counterparts\footnote{Eleven are determined to be blended via visual inspection (see Section~\ref{sec:otherdata}).} & 87 & \textquotedbl \\
\hspace{5mm} PACS 160$\,\mu$m counterparts & 44 & \textquotedbl \\
\hspace{5mm} X-ray counterparts\footnote{Detected in at least one X-ray band in \citet{luo17}.} & 46 & \ref{xraydata} \\
\\
\hline
\hline
\multicolumn{2}{l}{\bf{Summary of AGN Identification}} \\
\hline
\hline
Classification & Number & Section(s) \\
 \hline
X-ray AGN\footnote{Classified as AGN in \citet{luo17} via multiple criteria (see Section~\ref{sec:agn}).} & 31 & \ref{sec:xray}  \\
\hspace{5mm} $L_{\rm x,int}\geq$3\e{42} erg s$^{-1}$ & 23 & \ref{sec:xrayonly} \\
\hspace{5mm} Hard X-ray Spectrum & 11 & \textquotedbl  \\
\hspace{5mm} Optical Spectroscopy & 7 &  \ref{sec:opspec} \\
$L_{\rm x, int}/L_{6 \rm GHz}$ Excess & 38 & \ref{sec:xrayradio} \\
MIR Colors & 9 & \ref{sec:ircolors} \\
SED Fitting & 14\footnote{Including two tentative warm-excess candidates (see Section~\ref{sec:sedfitting}).} &  \ref{sec:sedfitting} \\
%\hspace{5mm} Warm Excess Candidate & &  \\
Radio Excess & 6 & \ref{sec:radio_ir} \\
Radio Flat Spectrum Source & 8 & \ref{sec:index} \\
\\
HLAGN\footnote{Moderate-to-high radiative luminosity AGN (HLAGN) and low-to-moderate radiative luminosity AGN (MLAGN) as defined in \citet{del17}.}   & 46 & \ref{sec:disc} \\
MLAGN & 3 & \textquotedbl\\
\hline
\label{tbl:summary}
\end{tabular}
\end{table}

\begin{table*}\footnotesize
%\begin{table}\pagewidthwidth
\centering
\caption{List of radio sources  in our 6 GHz parent sample at $z>0.75$ that satisfy one or more criteria for AGN identification.  A dash indicates the data were not available to assess the corresponding criteria for that source (not used for the optical spectroscopy column). In this table, $L_{\rm x, int}$ refers to the intrinsic 0.5-7 keV X-ray luminosity.} 
\begin{tabular}{cccccccccccc}
\hline
\hline
VLA & 3D-HST & $z$ & X-ray  & $L_{\rm x, int}\geq3\e{42}$ & Hard & $L_{\rm x, int}/L_{\rm 6 GHz}$ & MIR  & SED  & Radio & Radio & Optical \\
ID & ID &  & AGN\footnote{X-ray AGN as identified in \citet{luo17}. } & erg s$^{-1}$ & X-ray & Excess &  AGN &  AGN &  FSS\footnote{Flat Spectrum Source.} & Excess & Spectroscopy\footnote{NLAGN refers to high ionization narrow line AGN.  BLAGN refers to broad line AGN.  References: [1] \citet{san09}, [2] \citet{sil10}.} \\
\hline
%VLA033234.1-275029 & 15328 & 1.384 &  &  &  &  &  &  &  &  & \\
VLA033235.6-275021 & 15847 & 1.545 & - & - & - & - &  &  & x &  &  \\
%VLA033234.5-275005 & 16680 & 1.99666 &  &  &  &  &  &  &  &  &  \\
%VLA033237.7-275000 & 16960 & 1.4242 &  &  &  &  &  &  &  &  &  \\
VLA033231.8-274958 & 17069 & 0.99 &  &  &  & x &  &  & x &  & \\
VLA033244.5-274940 & 18006 & 1.016 & x & x & x & x &  &  &  &  & NLAGN [2] \\
VLA033248.6-274934 & 18251 & 1.115 & x &  & x & x &  &  &  & - &  \\
%VLA033248.5-274935 & 18398 & 1.12 & - & - & - & - &  &  &  & - &  \\
VLA033248.8-274936 & 18443 & 3.06 & - & - & - & - & x & - &   &  &  \\
%VLA033234.8-274921 & 19142 & 1.5979 & - & - & - & - &  & - & - &  &  \\
VLA033235.7-274916 & 19348 & 2.582 & x & x & x & x & x & x$\dagger$ &  &  & NLAGN [1] \\
VLA033243.5-274901 & 20137 & 1.508 & x &  &  & x &  &  &  &  &  \\
VLA033231.5-274854 & 20403 & 1.936 & x & x &  & x &  &  &  &  &  \\
%VLA033245.9-274855 & 20515 & 1.41935 & - & - & - & - &  &  & - &  &  \\
VLA033236.0-274850 & 20651 & 1.309 & x & x &  & x &  &  &  &  &  \\
VLA033239.7-274850 & 20788 & 3.064 & x & x & x & x & x &  &  &  & NLAGN [1,2] \\
VLA033243.7-274851 & 20808 & 2.50 & x & x &  & x & x & x$\dagger$ &  &  &  \\
VLA033243.0-274845 & 21205 & 1.730 & x & x & x & x &  &  &  &  &  \\
VLA033238.7-274840 & 21389 & 2.86 & x & x & x & x &  &  &  &  &  \\
VLA033244.6-274835 & 21615 & 2.593 & x & x &  & x &  & x &  &  &  \\
%VLA033234.8-274835 & 21740 & 1.245 & - & - & - & - &  &  &  &  &  \\
VLA033241.8-274825 & 22154 & 2.10 & - & - & - & - &  &  &  & x & \\
%VLA033236.4-274813 & 22766 & 1.4865 & - & - & - & - &  &  & - &  &  \\
%VLA033230.3-274758 & 23651 & 1.8085 & - & - & - & - &  &  & - & - &  \\
%VLA033231.1-274758 & 23684 & 2.7139 & - & - & - & - &  &  & - &  &  \\
%VLA033251.5-274758 & 23919 & 0.7809 & - & - & - & - &  &  &  &  & \\
VLA033240.1-274755 & 24110 & 1.998 &  &  &  & x &  &  &  &  & \\
VLA033240.3-274752 & 24193 & 3.13 & - & - & - & - &  & x &  & x &  \\
%VLA033250.5-274740 & 24846 & 2.2788 & - & - & - & - &  &  & - &  & \\
%VLA033242.7-274734 & 25452 & 1.427 & - & - & - & - &  &  &  &  & \\
%VLA033237.0-274727 & 25787 & 1.75936 & - & - & - & - &  &  &  &  &  \\
%VLA033236.2-274726 & 26272 & 1.50929 & - & - & - & - &  &  & - &  &  \\
VLA033235.8-274719 & 26550 & 1.912 &  &  &  & x &  &  & - &  &   \\
VLA033222.3-274711 & 26650 & 5.68 & - & - & - & - &  & - & x &  &  \\
VLA033239.6-274709 & 26915 & 1.317 & x &  &  & x &  &  & - &  & \\
%VLA033247.4-274711 & 27088 & 1.098 & - & - & - & - &  &  &  &  & \\
%VLA033249.5-274704 & 27306 & 2.7168 & - & - & - & - &  &  & - &  & \\
%VLA033242.4-274707 & 27565 & 1.31868 & - & - & - & - &  &  &  &  &  \\
%VLA033234.4-274659 & 27881 & 1.41331 & - & - & - & - &  &  &  &  &  \\
VLA033228.5-274658 & 27882 & 2.515 & x &  &  &  &  &  &  &  &  \\
VLA033243.6-274658 & 28022 & 1.566 &  &  &  & x &  &  &  &  &  \\
%VLA033241.6-274653 & 28090 & 1.4159 & - & - & - & - &  & - & - & - &  \\
VLA033232.5-274654 & 28190 & 1.441 &  &  &  & x &  &  & - &  &  \\
%VLA033241.7-274655 & 28319 & 1.77597 & - & - & - & - &  &  &  & - &   \\
%VLA033243.0-274650 & 28656 & 1.036 & - & - & - & - &  &  &  &  &  \\
VLA033243.3-274646 & 28723 & 2.66 & x & x &  &  & x & x & x & x &  \\
VLA033230.9-274649 & 28743 & 1.173 &  &  &  & x &  &  &  &  &  \\
VLA033235.1-274647 & 28844 & 2.497 & x &  &  & x &  &  & - &  &   \\
%VLA033234.9-274640 & 29155 & 1.097 & - & - & - & - &  &  &  &  &  \\
%VLA033235.7-274639 & 29349 & 1.5561 & - & - & - & - &  &  & - &  &  \\
%VLA033243.5-274639 & 29372 & 2.9139 & - & - & - & - & x &  &  &  &  \\
VLA033244.0-274635 & 29427 & 2.98 & x & x &  & x & x & x$\dagger$ &  &  &  \\
%VLA033250.9-274633 & 29484 & 1.3337 & - & - & - & - &  & - & - &  &  \\
VLA033238.5-274634 & 29606 & 2.543 & x & x &  & x &  & x &  & - &  \\
%VLA033241.0-274631 & 29643 & 2.5534 & - & - & - & - &  &  &  &  &  \\
VLA033236.6-274631 & 29730 & 0.999 &  &  &  & x &  &  & x & - &  \\
VLA033246.3-274632 & 29816 & 1.220 & x & x & x & x &  &  &  &  &  NLAGN [2] \\
%VLA033249.8-274629 & 29958 & 2.3445 & - & - & - & - &  &  &  &  &  \\
VLA033248.0-274626 & 29988 & 3.00 & - & - & - & - & x & x$\dagger$ &  &  &  \\
VLA033231.5-274623 & 30274 & 2.225 & x & x & x & x & x & x$\dagger$ &  &  & NLAGN [1] \\
%VLA033225.4-274617 & 30414 & 0.896 &  &  &  &  &  &  &  &  &  \\
%VLA033247.2-274620 & 30504 & 2.8 & - & - & - & - &  &  & - &  &  \\
VLA033239.7-274611 & 30534 & 1.546 & x & x &  & x &  &  &  &  &  \\
%VLA033252.5-274610 & 31041 & 2.6231 & - & - & - & - &  & - & - &  &  \\
VLA033223.6-274601 & 31240 & 1.033 & x &  & x & x &  &  &  &  &  \\
VLA033246.9-274605 & 31301 & 2.79 & - & - & - & - &  & x & - & - &  \\
%VLA033236.3-274600 & 31342 & 0.895 & - & - & - & - &  &  &  &  &  \\
VLA033233.0-274547 & 31425 & 0.947 & x & x &  & x &  &  &  &  &  \\
%VLA033239.3-274532 & 32021 & 1.09803 & - & - & - & - &  &  &  &  &  \\
%VLA033233.5-274547 & 32285 & 1.95205 & - & - & - & - &  &  & - & x &  \\
%VLA033253.0-274544 & 32509 & 1.325 & - & - & - & - &  &  &  &  &  \\
%VLA033231.3-274545 & 32513 & 1.56511 & - & - & - & - &  &  &  &  &  \\
VLA033252.3-274542 & 32577 & 1.355 &  &  &  & x &  &  &  &  &  \\
%VLA033232.9-274540 & 32690 & 1.96685 & - & - & - & - &  &  &  &  &  \\
%VLA033241.3-274538 & 32701 & 1.54594 & - & - & - & - &  &  &  &  &  \\
VLA033230.0-274530 & 32932 & 1.221 & x & x &  & x &  & x &  &  &  BLAGN [1,2] \\
VLA033229.9-274521 & 33103 & 0.954 &  &  &  & x &  & - & - &  &  \\
VLA033230.1-274523 & 33287 & 0.955 & x & x &  & x &  & x &  & - & BLAGN [1] \\
%VLA033247.2-274525 & 33432 & 4.5471 & - & - & - & - &  & - &  &  &  \\
%VLA033246.9-274513 & 33849 & 0.84227 & - & - & - & - &  &  &  &  &  \\
%VLA033228.3-274519 & 33859 & 1.79128 & - & - & - & - &  &  & - &  &  \\
VLA033229.2-274510 & 34114 & 2.59 & x & x &  & x &  &  &  &  &  \\
%VLA033233.7-274508 & 34298 & 1.50052 & - & - & - & - &  &  & - &  &  \\
%VLA033237.7-274505 & 34518 & 1.21333 & - & - & - & - &  &  & - &  &  \\
%VLA033241.4-274458 & 34802 & 1.202 & - & - & - & - &  &  &  &  &  \\
VLA033247.6-274452 & 35127 & 1.569 & x &  &  & x &  &  &  &  &  \\
VLA033238.1-274432 & 35774 & 1.221 & - & - & - & - &  &  &  & x &  \\
%VLA033247.9-274433 & 35951 & 2.14873 & - & - & - & - &  &  &  &  &  \\
VLA033228.8-274435 & 36223 & 5.50 & x & x &  &  &  &  & x &  &  \\
VLA033241.0-274427 & 36451 & 1.302 & x & x & x & x &  &  &  & x &  \\
VLA033236.0-274424 & 36653 & 1.038 &  &  &  & x &  &  & - &  &  \\
%VLA033237.1-274419 & 37005 & 3.0672 & - & - & - & - &  &  &  &  &  \\
%VLA033246.3-274418 & 37124 & 2.4345 & - & - & - & - &  &  &  &  &  \\
VLA033235.1-274410 & 37387 & 0.839 & x &  &  & x &  & x & x &  &  \\
VLA033242.7-274407 & 37620 & 3.14 & - & - & - & - &  & x & x &  &  \\
VLA033229.8-274400 & 37885 & 2.08 & - & - & - & - &  &  & - & x &  \\
VLA033238.0-274400 & 37989 & 3.30 & x & x &  & x & x & - &  &  &  \\
%VLA033232.1-274355 & 38269 & 2.5712 & - & - & - & - &  &  &  &  &  \\
%VLA033242.7-274348 & 38704 & 2.4607 & - & - & - & - &  &  &  &  &  \\
VLA033241.7-274328 & 39751 & 0.979 & x & x & x & x &  &  &  &  &  \\
% 51 & & & 31 & 23 & 11 & 38 & 9 & 12 (14) & 8 & 6 &  7 \\
\hline
\multicolumn{4}{l}{$\dagger$ \textrm{Warm-excess AGN candidate (Section~\ref{sec:sedfitting}).}}
\label{tbl:simple}
\end{tabular}%
\end{table*}

\subsection{X-ray Identified AGN} \label{sec:xray}

X-ray (and optical) based source classifications (galaxy or AGN) are adopted and expanded upon following the criteria from \citet{luo17}: {\bf (1)} X-ray properties only, i.e. a high intrinsic X-ray luminosity and/or a hard X-ray spectrum, {\bf (2)} excess X-ray relative to other multi-wavelength properties, i.e. a high ratio of X-ray to the R-band, K$_{s}$-band, or radio flux, and {\bf (3)} optical spectral features (see Section~\ref{sec:opspec}). 

\subsubsection{Selection via X-ray Properties Only}\label{sec:xrayonly}

\begin{figure*}[ht!]
\centering
\includegraphics[width=0.8\textwidth]{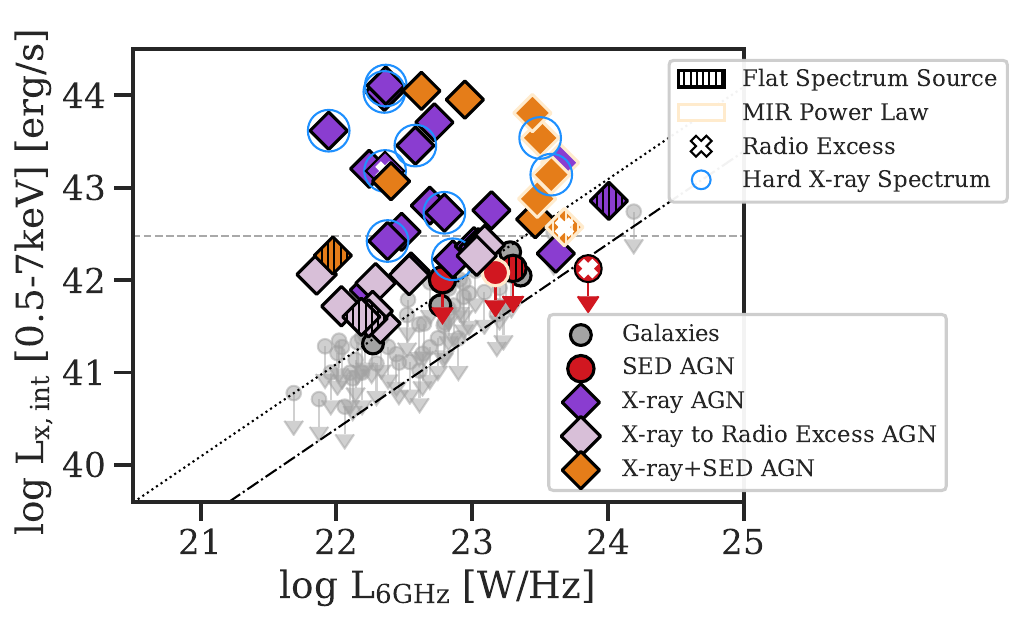}
\caption{The intrinsic 0.5-7 keV X-ray luminosity as a function of 6 GHz luminosity (assuming $\alpha=-0.7$) for our 100 radio sources. Symbols represent AGN identified primarily via X-ray properties (dark purple diamonds), SED fitting (red circles), or both (orange diamonds), with non-AGN galaxies shown as small gray circles.  New AGN identified via their X-ray to 6 GHz excess are depicted in light purple (Section~\ref{sec:xrayradio}). Upper limits are $3\sigma$.  Additional AGN indicators are shown as follows: MIR power law emission is outlined in beige (Section~\ref{sec:ircolors}), outliers from the radio-infrared correlation are marked with white x's (Section~\ref{sec:radio_ir}), and flat spectrum radio sources are denoted by vertical hatch (Section~\ref{sec:index}).  Hard X-ray sources are indicated with blue circles.
The dot-dash line denotes the nominal $L_{x}-L_{\rm radio}$ threshold expected for a maximal starburst.  We adopt a conservative cut at 5x this level (dotted line) for the identification of AGN via excess X-ray relative to radio emission. 
}
\label{fig:xray}
\end{figure*}

Figure~\ref{fig:xray} shows the intrinsic (absorption corrected, see \citet{luo17}) X-ray luminosities of our sample as a function of $L_{\rm 6 GHz}$, with hard X-ray sources highlighted.  
Of our 100 6 GHz radio sources, 46 are detected in the \citet{luo17} 7 Ms catalog, with 32 classified as X-ray AGN by that work, although one source
%, 18398, 
is likely to be a mis-identification resulting from blending in the X-ray and is excluded from our AGN list. The majority of these AGN are identified primarily via high intrinsic X-ray luminosity (see \citet{luo17} for details).  The threshold for this luminosity cut varies slightly within the literature, typically around $\sim10^{42}$ erg s$^{-1}$, corresponding to the X-ray emission expected from the most luminous star forming galaxies \citep[i.e.][]{ale05}.  In \citet{luo17}, a conservative threshold of $L_{0.5-7 keV} \geq 3\e{42}$ erg s$^{-1}$ is adopted \citep[see also][]{xue11, leh16}.   AGN meeting this criterion are shown in Table~\ref{tbl:simple}.   

Obscured AGN can be identified via a hard X-ray spectrum \citep[see][for a review]{bra15}.  Both hard and soft band detections are available for about half of the X-ray detections for our radio sample.  For these sources, \citet{luo17} measured the effective photon index,  $\Gamma_{\rm eff}$.  The range for our radio sample is $-1.5<\Gamma_{\rm eff}<2.5$, with 11 of our X-ray AGN having $\Gamma_{\rm eff}<1.0$, indicative of an obscured AGN (see Table~\ref{tbl:simple}).

\subsubsection{Excess X-ray Relative to Other Properties} \label{sec:xrayradio}

AGN can also be identified through  excess X-ray emission relative to the output in the optical, near-infrared, and/or radio.  As seen in Table~\ref{tbl:simple}, \citet{luo17} identified 6/31 X-ray AGN through their excess in the X-ray relative to other bands, rather than through high X-ray luminosities or hardness ratios. Here we expand on the excess X-ray to radio emission selection using our 6 GHz data, $\sim9 \times$ deeper than the 1.4 GHz data employed in \citet{luo17}.

Using the ratio of X-ray to radio for AGN identification hinges on the question of what powers the radio emission.  In inactive star forming galaxies, we can expect a relation between X-ray and radio emission, as both are dominated by mechanisms related to young star formation \citep[e.g.,][]{vat12}.  In the presence of an AGN, however,  X-ray and radio emission can be sensitive to very different physical processes. The X-ray spectrum  is dominated by hot coronal gas near the BH \citep{bra15}, a ubiquitous feature in luminous AGN, modulo heavy obscuration.  Radio emission, conversely, can be generated from multiple sources associated with the AGN including e.g. large or small-scale jets \citep{gal06}, outflows and/or winds \citep{blu01, kin13, zak14, nim15}, or electron acceleration within the hot corona \citep{lao08, rag16}. If these mechanisms are weak or non-existent, then radio can trace star formation in the host galaxy while X-ray is dominated by the AGN (see Section~\ref{sec:hostgals} for a detailed discussion on what is powering the radio in our sample).  

The $L_{x}-L_{\rm radio}$ relation for SFGs has been established empirically both locally \citep[i.e. ][]{ran03, ran12, per07, leh10} and at higher redshift \citep[i.e.][]{vat12} with consistent results across a range of luminosities and little evidence for redshift evolution \citep[but see ][]{leh16}.  There is, however, significant ($\sim0.4$ dex) intrinsic scatter in the $L_x-$SFR relation \citep{min14,sym14,leh16} as well as the radio-infrared correlation \citep[0.25 dex for $q_{24}$;][]{yun01,rie09}, which will propagate here.  Given the $L_{x}-L_{\rm radio}$ relation  for SFGs and the maximal X-ray emission expected for the most luminous SFGs \citep{ale05}, \citet{xue11} and \citet{leh16} derived a threshold for excess X-ray over radio due to AGN activity as $L_{0.5-7 \rm keV}/L_{\rm 1.4 GHz} \geq \beta \times 8.9\e{17}$ where $\beta$ is the level of excess.  To be conservative given population scatter, we adopt $\beta=5$ and convert to $L_{6 \rm GHz}$ using $\alpha=-0.7$, yielding $L_{0.5-7 keV}/L_{6 \rm GHz} \geq 1.2\e{19}$ as our X-ray to radio excess criterion.

The X-ray to radio excess sources are indicated in Table~\ref{tbl:simple}.  38 AGN are identified via this method, with 10 new AGN candidates based on this criterion displayed as light purple diamonds in Figure~\ref{fig:xray}.  

\subsection{Optical Spectroscopy}\label{sec:opspec}

Optical spectroscopy is one of the criteria used to identify AGN in the \citet{luo17} source classifications, adopted in this work.  The classification of AGN via optical spectroscopy falls into two categories: broad emission lines, which indicate a Type-1, unobscured AGN and highly ionized narrow emission lines, which identify Type-2, obscured AGN.  Optically identified AGN in GOODS-S over the relevant redshift range were also compiled and categorized in \citet{san09} and \citet{sil10} from the various references therein.  Matching these to our radio sample, we find our optically-selected AGN can be categorized as 2 broad line AGN (BLAGN) and 5 narrow line AGN (NLAGN), identified in Table~\ref{tbl:simple}.

\subsection{Selection in the Mid-IR} \label{sec:mir}

MIR colors have been shown to be effective at selecting luminous AGN, which produce a distinctive power law (PL) spectrum in this spectral range \citep{lac04, ste05, don12, kir13}. Some of these sources can be heavily obscured and are often missed in X-ray surveys \citep{don12,del16,  del17}.  Here we utilize two techniques to identify these AGN: selection in IRAC color-color space and SED fitting using the full optical-MIR photometry available from 3D-HST.

\subsubsection{Selection by IRAC Colors} \label{sec:ircolors}

\begin{figure}[ht!]
\includegraphics[scale=0.55, trim=2mm 0 0 0, clip]{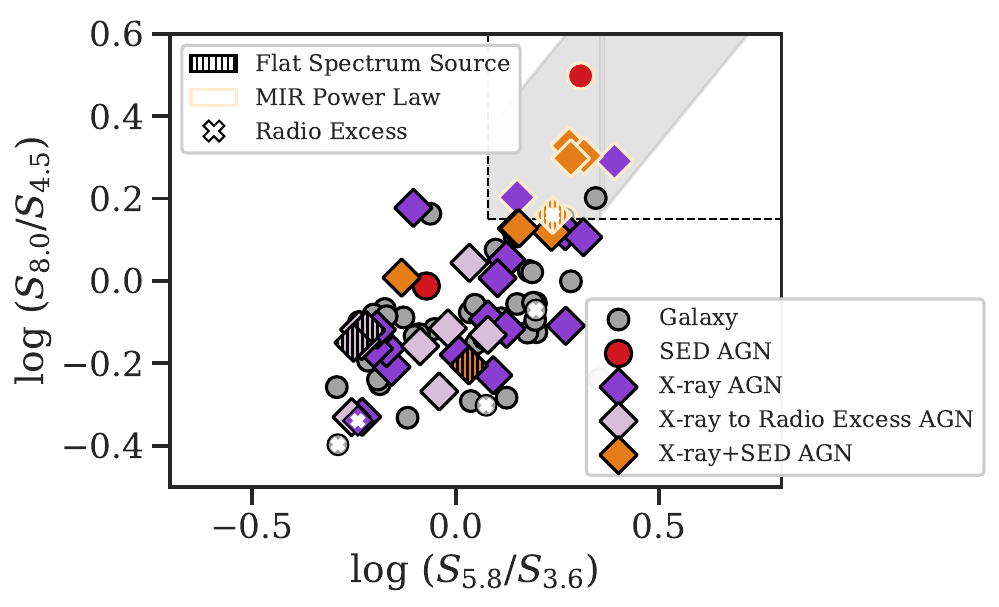}
\caption{The {\it Spitzer}/IRAC MIR colors of our radio sample, identifying power law AGN.  Symbols are as in Figure~\ref{fig:xray}.  For this selection, we require a $>3\sigma$ detection in all four IRAC bands and $z<4$ (see Section~\ref{sec:ircolors}). We adopt the color cuts (dashed lines) from \citet{kir13}, designed to avoid contamination from SFGs.  The shaded region denotes a similar selection from \citet{don12}. One radio source  (gray circles) within the shaded region is discarded as an AGN candidate due to lack of indication of PL emission in the full SED.  
}
\label{fig:iraccolors}
\end{figure}

In Figure~\ref{fig:iraccolors}, we show the MIR colors of our radio sources, based on the {\it Spitzer} IRAC (3.6, 4.5, 5.8, 8.0$\,\mu$m) bands  \citep{lac04,ste05, alo06}.  13/100 of our sources are not considered for this criterion as they do not have $>3\sigma$ measurements in all four IRAC bands and/or are at $z>4$\footnote{At $z>4$, the IRAC bands sample stellar emission short of $1.6\,\mu$m, which can mimic an AGN power law. }.

We find 10 power law (PL) AGN candidates following the criteria outlined in \citet{kir13}, designed to minimize contamination by SFGs \citep[see also][]{don12}.  Seven of these PL AGN are also identified via other selections.  Of the remaining three, one is firmly a MIR PL AGN, but lacks an X-ray detection or adequate photometry for SED fitting; a second is a candidate for a warm-excess AGN (see the next section); and the third is a marginal MIR color candidate with no indication of PL emission in a visual inspection of the SED.  We discard the third candidate in all subsequent analysis. The MIR PL AGN are indicated in Table~\ref{tbl:simple}.

\subsubsection{SED Fitting and Decomposition} \label{sec:sedfitting}

With the extensive photometry available for most of our sources, SED fitting provides a more diagnostic approach than just using photometric colors. 
In this section, we describe our  fitting to identify AGN that are apparent above the optical-MIR emission of the host galaxy. In addition, SED decomposition puts constraints on the fractional contribution of the AGN to the galaxy's optical-MIR emission.

For SED fitting, we require that at least 9 optical-MIR photometric bands have SNR$>3$ and {\tt use\_phot}=1 in the 3D-HST catalog, indicating a robust photometric measurement.  These criteria are satisfied for 91/100 of the 6\,GHz radio sample.  We utilize the publicly available SED fitting code from \citet{assef10}, with modifications, to perform AGN identification and SED decomposition over the rest wavelength range 0.03 $-$ 30$\,\mu$m.  This code performs a non-negative linear combination of templates, applying galaxy templates first and then galaxy+AGN templates. A detailed analysis of how this method compares to other common AGN indicators was presented in \citet{chu14}.  In summary, they found that this SED fitting method identifies a population of AGN that only partially overlaps with X-ray selection and correlates well with AGN selected via optical spectroscopy.  In general, this method will identify AGN that have moderate to strong contributions to optical-MIR emission independent of X-ray obscuration, but will fail to identify AGN where the host galaxy emission dominates (see also SED3FIT from \citet{ber13} and subsequent studies for similar analysis).

Our template set consists of 16 star forming spectra with variable optical attenuation ($A_v=0-3$), representing young stellar populations; the \citet{assef10} ``elliptical'' template, representing emission from the old stellar population; and, for the AGN contribution, we adopt a standard Type-1 AGN template from \citet{elv94} with host galaxy contribution removed \citep{xu15}. More details of the galaxy and AGN templates used and the fitting procedure are provided in Appendices~\ref{app:modeling} and \ref{app:fsps}.   A fit with an AGN component is preferred if it is a significant improvement over a galaxy-only fit using the Fisher test, which evaluates whether an additional component or free parameter can improve a fit by chance (via the F-ratio, see Appendix~\ref{app:modeling}).  From this test, we calculate the purity (F$_{\rm prob}$) of a sample with F-ratio larger than a given number, accounting for the relative rarity of AGN, as described in greater detail in \citet{chu14}.  We adopt a threshold of F$_{\rm prob}\geq0.93$ for AGN identification. Following the fitting procedure, a visual inspection is performed on all fits.

\begin{figure*}[h!]
\includegraphics[width=0.5\linewidth, trim=0 0 0 0, clip]{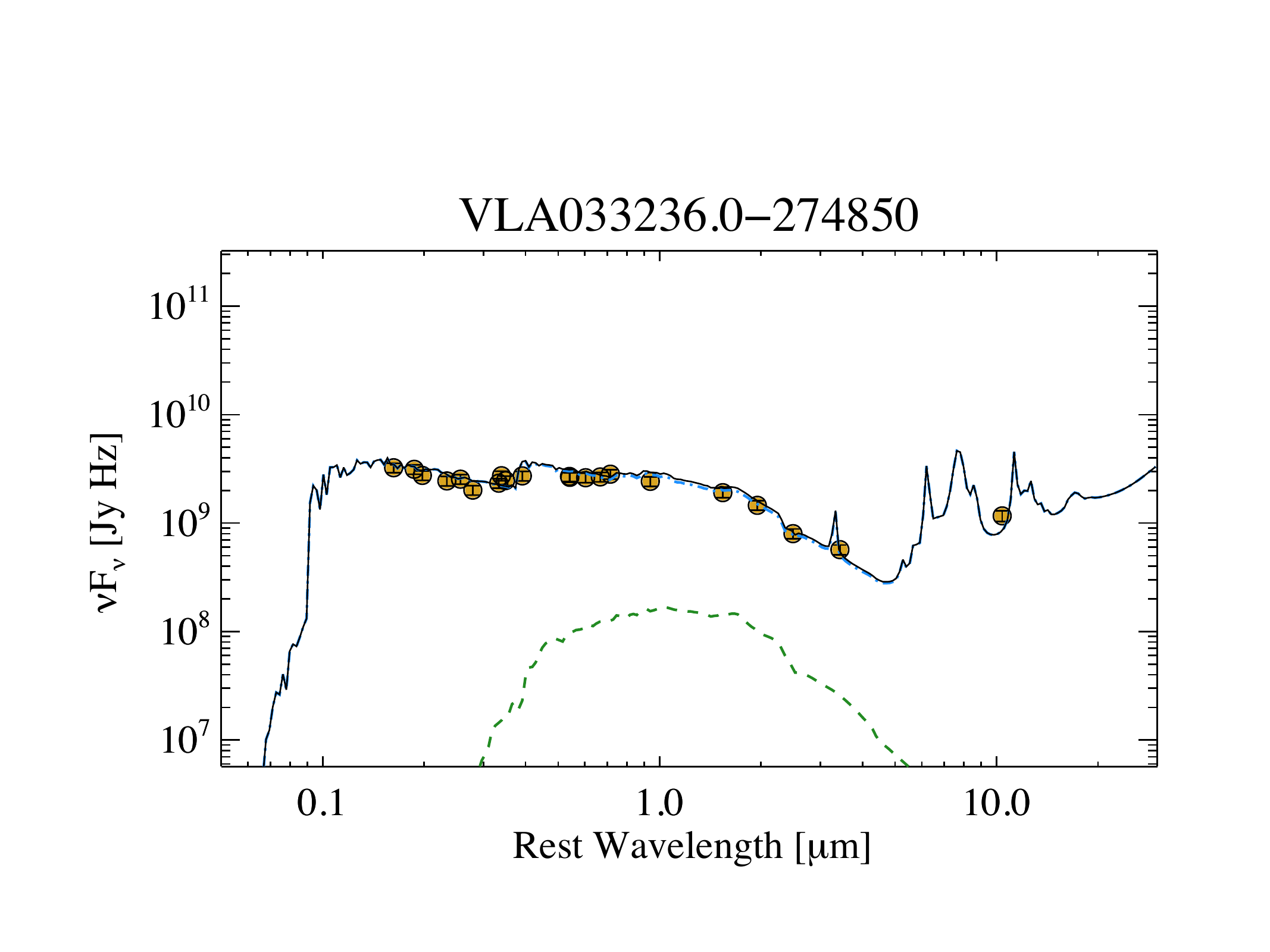}
\includegraphics[width=0.5\linewidth, trim=15mm 15mm 0 0, clip]{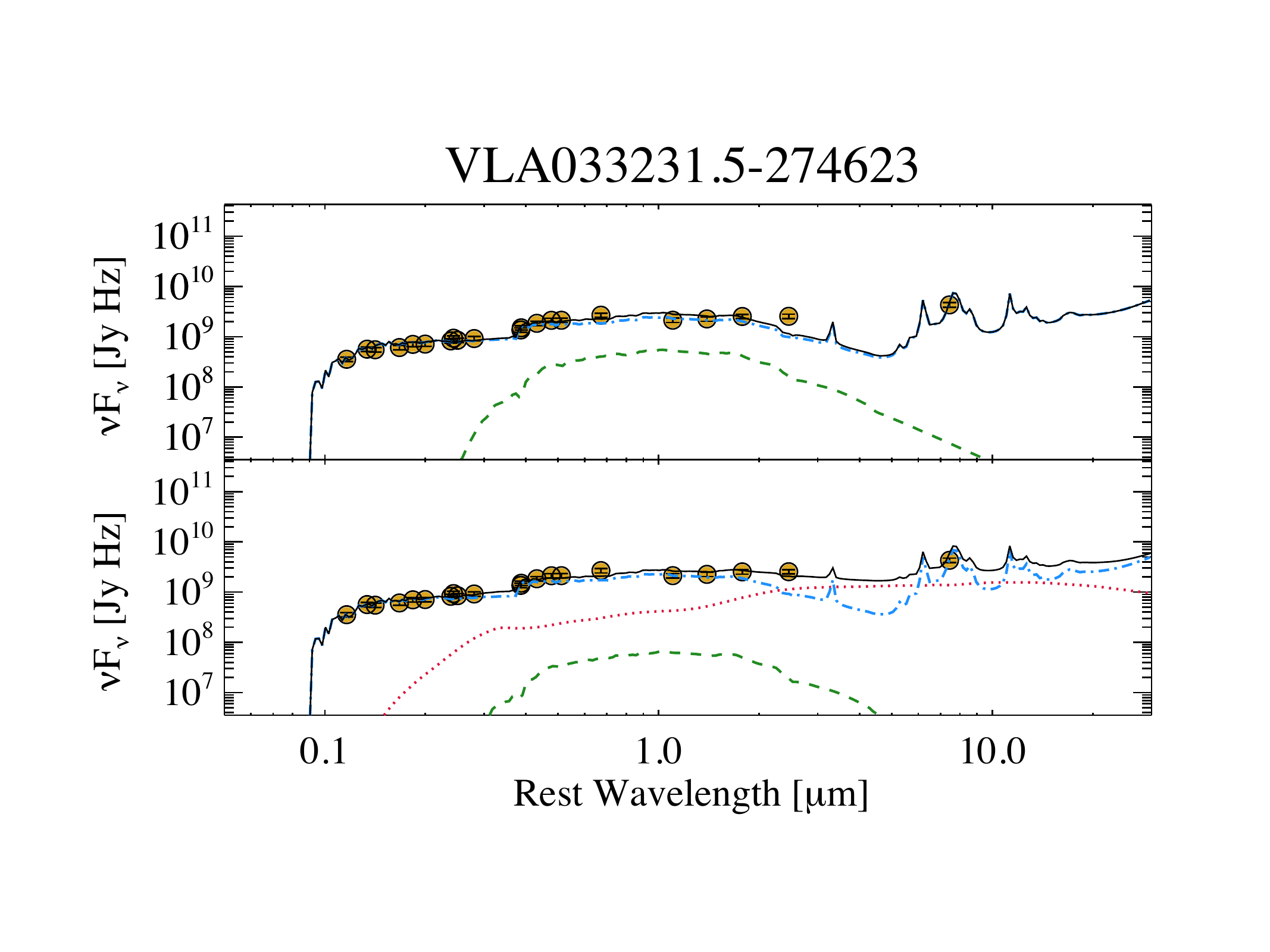}
\caption{{\it (left)} An example of the SED fit of a star forming galaxy at $z_{spec}=1.309$.  The solid black line is the total fit SED, composed of a SFG template with A$_{\rm v}=0.6$ (dot dash, blue), and an old stellar population template (dashed, green), fit to the observed datapoints (yellow circles).  {\it (right)} A comparison between a fit with galaxy templates only (top) and with a standard Type-1 AGN template (dotted, red) added (bottom) for a radio source at $z_{spec}=2.225$.  The AGN template fills in the MIR excess in the region of the stellar minimum at rest $\sim3-5\,\mu$m. The possibility that the AGN improves the fit only by chance is ruled out at the 98$\%$ level.}
\label{fig:fitex}
\end{figure*}

\begin{figure*}[h!]
\includegraphics[width=0.5\linewidth,trim=0 40mm 0 0, clip]{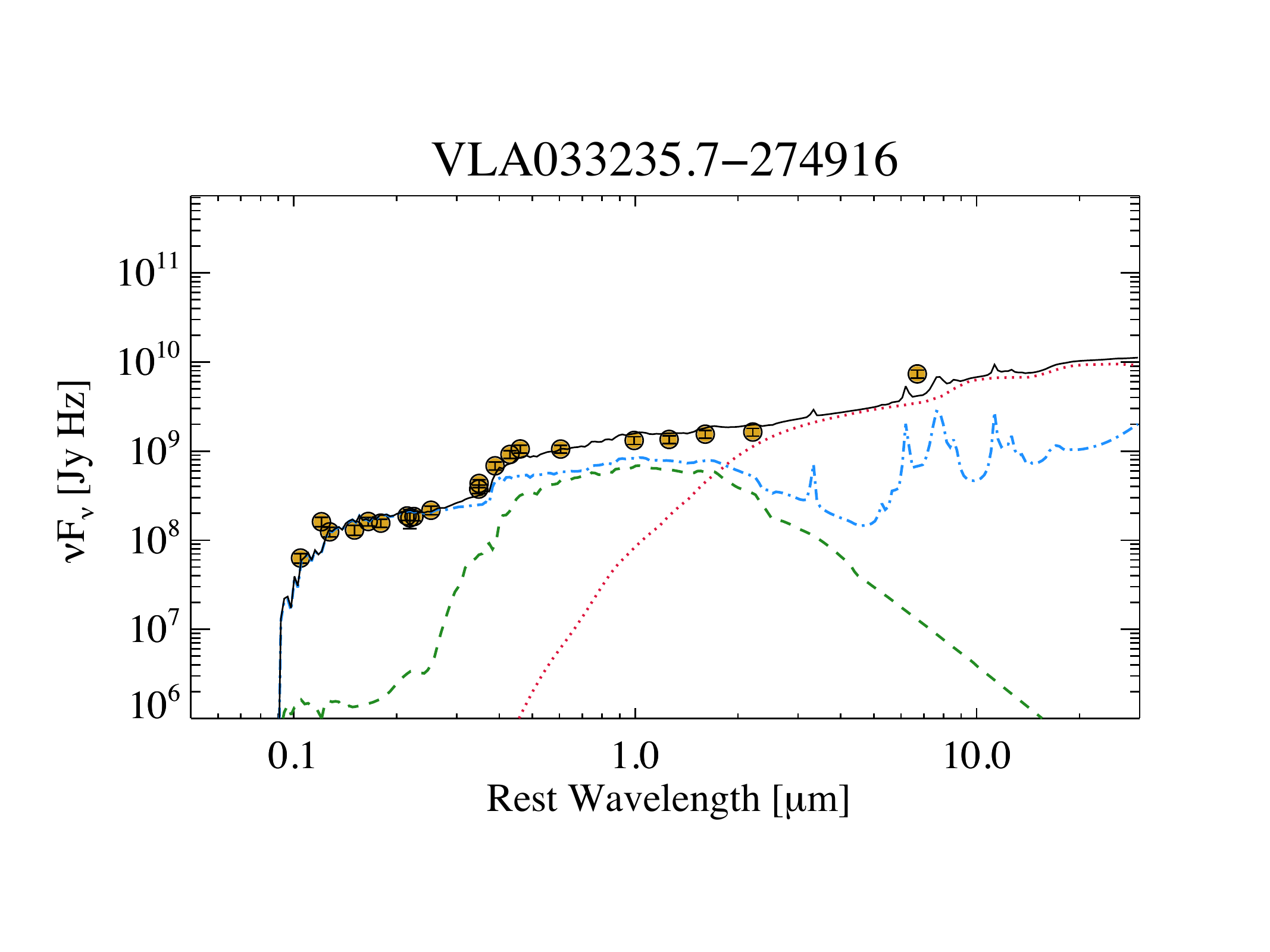}
\includegraphics[width=0.5\linewidth,trim=0 40mm 0 0, clip]{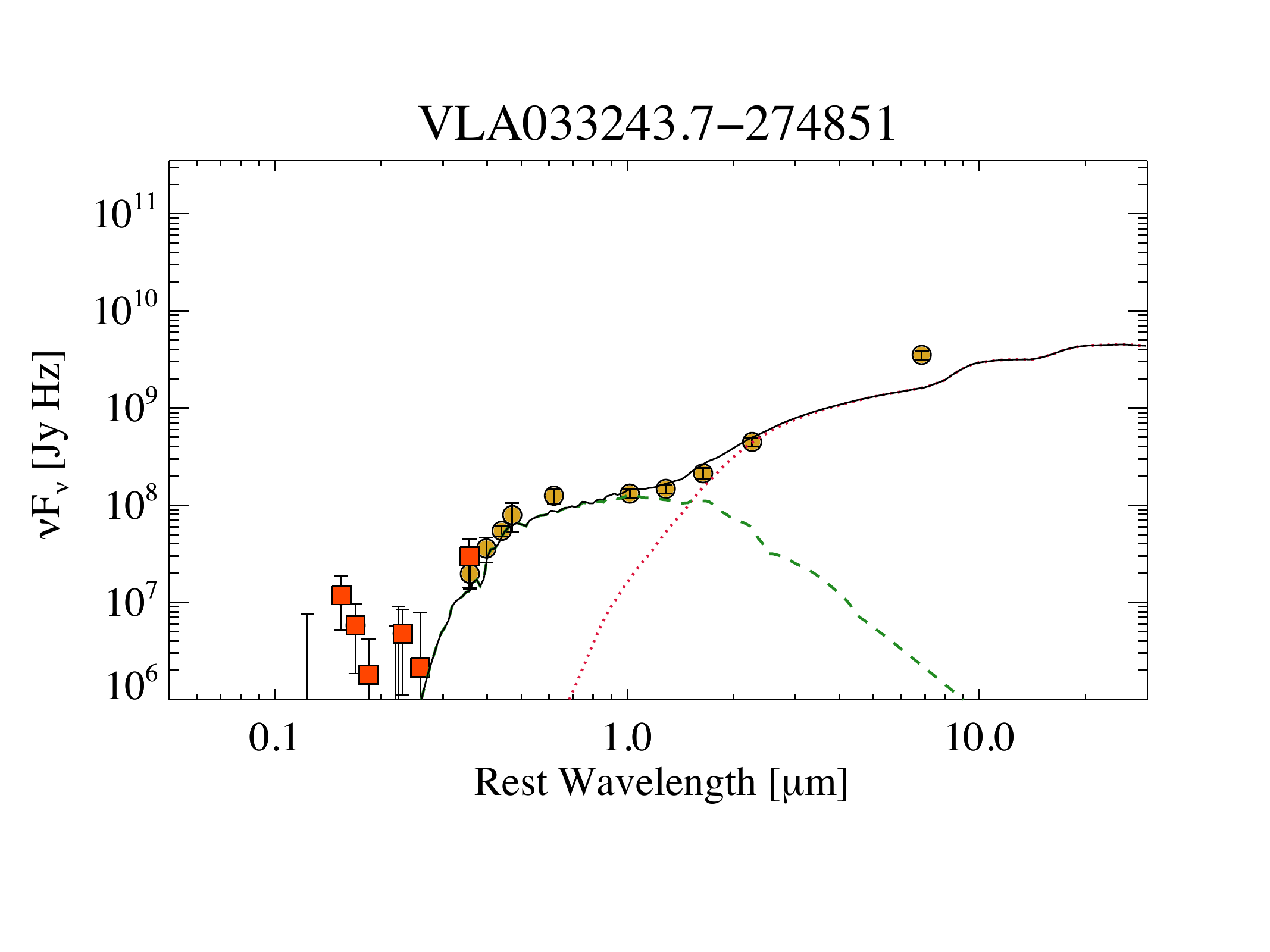}
\caption{Two examples of MIR warm-excess AGN at $z\sim2.5$ that are best fit with an additional obscured polar dust component, as described in \citet{lyu18}.  Symbols and lines are as in Figure~\ref{fig:fitex}, with red squares indicating data below the detection limit that is not included in the fit. The reddening and extra emission by polar dust results in a significantly different AGN SED, as can be seen by comparing with Figure~\ref{fig:fitex} (right, bottom). }
\label{fig:badfits}
\end{figure*}

Figure~\ref{fig:fitex} (left) shows the fit for VLA033236.0-274850, indicated to be purely star forming. Figure~\ref{fig:fitex} (right) displays radio source VLA033231.5-274623, fit with galaxy templates only (top) and with galaxy+AGN templates (bottom).  This latter source has F$_{\rm prob}=0.98$ $-$ in other words, the probability that adding the AGN template improved the fit {\bf only} by chance is $2\%$ $-$ so this source is identified as having an AGN via SED fitting. This particular example is also identified as an AGN via its X-ray properties. 

SED fitting for AGN identification such as the procedure described above is limited by the currently available photometric coverage. The intrinsic spectrum of the majority of Type-1 AGN can be robustly described by the single AGN template used in these fits, empirically derived originally in \citet{elv94}, with modifications to remove the remaining host contamination \citep{xu15}.  Recent works have begun to quantify deviations from this Type-1 spectrum, revealing sub-populations of hot- and warm-dust deficient quasars \citep[i.e.][]{lyu17} as well as  contributions from line-of-sight polar dust \citep{lyu18}.  Unfortunately, the constraining power of current photometric datasets to distinguish these different cases in high redshift populations is limited, due to sparse coverage of the infrared spectrum.  At the typical redshifts of our sources, our long wavelength coverage consists of the IRAC and MIPS 24$\,\mu$m bands, missing the crucial regions containing the stellar minimum at rest $3-5\,\mu$m and most of the $6-12\,\mu$m region, which is dominated by aromatic bands in SFGs and can probe the presence of warm dust in AGN \citep{lyu17}. A more detailed analysis in these cases requires additional constraints in the infrared. Similarly, these constraints will be vital to the inclusion of Type-2 templates, which we have not included here due to the lack of convergence on simple forms of the SED suitable for our fitting procedure and available datapoints.  For statistical samples of sources, constraining power will be provided by the $5-25.5\,\mu$m photometric coverage of {\it JWST's} Mid-Infrared Instrument (MIRI) up to $z\sim2.5$ (see Sections~\ref{sec:census}-\ref{sec:future} for further discussion).  

Despite these limitations, we do see some indication of variety in the AGN SEDs within our sample.  In five AGN with adequate photometric coverage for SED fitting, we find evidence of warm MIR excess, requiring a steeper MIR slope than can be provided by the available combinations of galaxy and the \citet{elv94} template.  Such warm-excess sources have been noted in multiple AGN populations, from ``normal" AGN \citep{kir15, lyu17} to extremely red quasars \citep{ros15,ham17} and hot dust obscured galaxies \citep{eis12, wu12}.  As demonstrated in \citet{lyu18}, these populations can be modeled with the addition of an extended polar dust component, which adds obscuration along the line-of-sight to otherwise unobscured Type-1 AGN and pumps up the MIR.  To test whether our warm-excess sources are better described with a polar dust component, we adopt a template from \citet{lyu18} that combines the \citet{elv94} template with a polar dust component with $\tau_{\nu}=3$, and redo our fits\footnote{We note that \citet{lyu18} found that the polar dust SED is not very sensitive to the optical depth at moderate values ($\tau_{\nu}<5$).}.    We find that 4/5 of the warm-excess candidates are better fit with the \citet{lyu18} template, with the last source degenerate between the \citet{elv94} and \citet{lyu18} templates.  Of the four, two  (VLA033235.7-274916 and VLA033243.7-274851), shown in Figure \ref{fig:badfits}, meet our threshold for an AGN (F$_{\rm prob}>0.93$) while the other two (VLA033244.0-274635 and VLA033248.0-274626) fall just below at F$_{\rm prob}$=0.87 and 0.88. Following the inclusion of the \citet{lyu18} template, the final median reduced $\chi^2$ of all our fits is 2.7.

Including the two warm-excess AGN that meet our threshold, we identify 12 AGN in our radio sample via SED fitting, with an additional two tentative warm-excess candidates.  One of these tentative warm-excess candidates (VLA033248.0-274626) was identified in Section~\ref{sec:ircolors} via MIR colors, the other (VLA033244.0-274635) is X-ray identified.  This gives us confidence that these candidates are real AGN and we adopt the best-fit \citet{lyu18} template when deriving AGN parameters for all the warm-excess candidates.  One additional AGN was robustly identified via MIR colors only (VLA033248.8-274936), lacking  both an X-ray detection and adequate photometry to do SED fitting, bringing the total number of MIR AGN to 15.  Of these, only 5 were not previously identified via their X-ray properties.

This high degree of overlap in X-ray and MIR selections can be compared to the overlap in a similar radio-based study from \citet{del17}.  They found that only about a third of their MIR AGN are also X-ray AGN, whereas we find 10/15 of our MIR AGN are also identified via the X-ray.  We attribute this difference to the high completeness of the 7 Ms Chandra catalog down to the lowest X-ray luminosities associated with unobscured or moderately obscured AGN ($10^{42}<L_x<10^{44}$ erg s$^{-1}$).

\begin{figure}[ht!]
\includegraphics[scale=0.64, trim=2mm 0 4mm 0, clip]{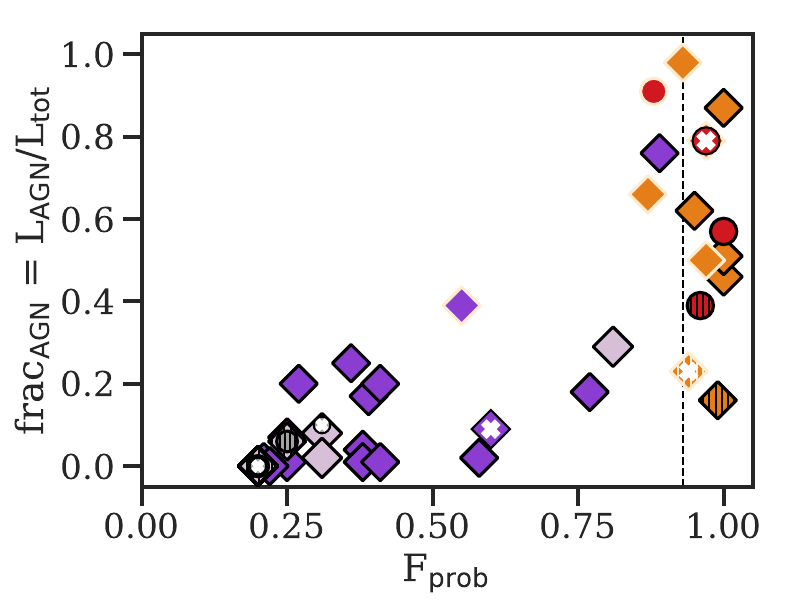}
\caption{The fractional contribution of the AGN emission to the total optical-MIR SED derived via SED fitting, frac$_{\rm AGN}$, as a function the F-test probability, F$_{\rm prob}$, that the galaxy hosts an AGN.  Symbols are as in Figure~\ref{fig:xray}. AGN identified via SED fitting have a range of AGN to host contributions while the majority of AGN identified primarily via X-ray properties have $<25\%$ of their optical-MIR SED supplied by the AGN.  The vertical dashed line denotes our selection threshold for SED-identified AGN (red circles and orange triangles); the two SED-identified AGN that fall below this threshold are warm-excess candidates.  }
\label{fig:fprob}
\end{figure}

The SED fits let us compare the AGN and host galaxy luminosities. We calculate the 0.03-30$\,\mu$m bolometric luminosities for the AGN and galaxy components, and evaluate the AGN contribution via frac$_{\rm AGN} = L_{\rm 0.03-30\mu m}^{\rm AGN} / L_{\rm 0.03-30\mu m}^{\rm total}$, where frac$_{\rm AGN}=1$ is completely AGN dominated.  Figure~\ref{fig:fprob} shows the fractional AGN contribution as a function of the F-test probability that the source has an AGN, where the galaxy+AGN fit results have been adopted for all AGN identified via any method. AGN identified via SED-fitting have a range of frac$_{\rm AGN}$ values, indicating that this selection method is sensitive to both host- and AGN-dominated optical-MIR SEDs.  The majority of AGN in our radio sample, however, are identified via X-ray properties. AGN not indicated via MIR selections are primarily host-dominated, with $<25\%$ of their optical-MIR emission contributed by the AGN.  For one quarter of our full AGN sample, galaxy+AGN fits return no AGN contribution (frac$_{\rm AGN}=0$). This result emphasizes the complementarity of the different selection techniques and shows the importance of X-ray and radio identification in finding lower luminosity AGN that are overwhelmed by their host galaxies in the MIR.

\subsection{Identification of AGN via Radio Properties} \label{sec:indent}

There are several metrics for identifying AGN via radio properties.  In the following sections, we discuss these radio indicators of AGN activity: radio luminosity, radio morphology, outliers from the radio-infrared correlation, and flat or inverted radio spectral slopes.

\subsubsection{Radio Luminosity and Morphology}

Radio-loud or jetted AGN \citep{pad16} are dominated by synchrotron emission associated with large-scale extragalactic jets.  These AGN are traditionally identified via a luminosity cut, with AGN labeled RL above $L_{1.4 \rm GHz}\gtrsim10^{25}$ W Hz$^{-1}$ \citep[$L_{6 \rm GHz}\gtrsim 10^{24.6}$ W Hz$^{-1}$;][]{mil90}.  Morphological signatures can also identify RL AGN, as kpc-scale extragalactic jets can be easily recognized in radio images with sufficient resolution.  Neither a straight luminosity cut nor visual morphological selection are considered complete selections, however, and so it is common in the literature to also examine the radio emission in relation to other portions of the SED; for example, the X-ray to radio ratio \citep{ter03}, optical to radio ratio \citep{kel98}, and the radio-infrared correlation \citep{hel85, con91, yun01}. 

Our radio sample falls entirely below the traditional RL AGN luminosity cut (assuming $\alpha=-0.7$; Figure~\ref{fig:radio}).  This is consistent with a visual inspection of our sub-arcsecond resolution 6 GHz radio image, which reveals no evidence of large-scale extragalactic jets.  These base-level checks, together with the fact that RL AGN are relatively rare \citep[$\lesssim10-20\%$;][]{bon13, wil15}, suggest that there should be few or even no RL AGN in our sample.  In the next section, we expand our search for RL AGN via the radio-infrared correlation, using the q$_{24}$ parameter \citep{don05, bon13}. These AGN may be missed by non-radio indicators due to the AGN being in a phase of less efficient BH accretion \citep{del17}.

\subsubsection{The Radio-IR Correlation}\label{sec:radio_ir}

The goal of this section is to identify RL AGN via outliers from the well-established radio-infrared correlation for SFGs \citep{hel85, con91, yun01}, which holds over a wide range of luminosities and has a weak, if any, dependence on redshift \citep{ivi10, mao11, mag15, delh17}.  This relation is typically evaluated via the q-parameter, the ratio of infrared, via mid-, far-, or total infrared emission, to the 1.4 GHz radio emission.  To match the depth of our radio observations in the infrared as well as possible, we adopt q$_{24,\rm obs}$ = log($S_{24,\rm obs}/S_{1.4 \rm GHz,obs}$), where $S_{24,\rm obs}$ is the observed MIPS 24$\,\mu$m flux density and $S_{1.4 \rm GHz,obs}$ is the 1.4 GHz flux density \citep{app04, don05, bon13}.  We use q$_{24}$ rather than q$_{160}$, q$_{250}$, or far infrared luminosity because far-infrared counterparts are only detected for approximately half our radio sample; these are utilized later in the discussion to explore the origin of radio emission (see Section~\ref{sec:q160}).

Deriving the 1.4 GHz flux density requires that we extrapolate from 6 GHz, under the assumption that the 1.4-6 GHz radio spectrum can be described as a power law.  Recent works by \citet{delh17} and \citet{tis19} evaluated the sensitivity of the SFG locus of the radio-IR correlation to the assumed radio spectral slope, $\alpha$.  They found that, though different assumptions for $\alpha$ can affect the normalization and redshift dependence of the SFG locus, the effect is small \citep[at the $\sim20\%$ level for the derived luminosity;][]{tis19} in comparison to the intrinsic scatter in the relation.  Therefore, we assume the fiducial value of $\alpha=-0.7$ when deriving the 1.4 GHz flux density from 6 GHz, with the caveat that we may mis-classify sources with extreme slopes.

\begin{figure}[ht!]
\hspace{-5mm}
\includegraphics[scale=0.68, trim=3mm 0 4mm 0, clip]{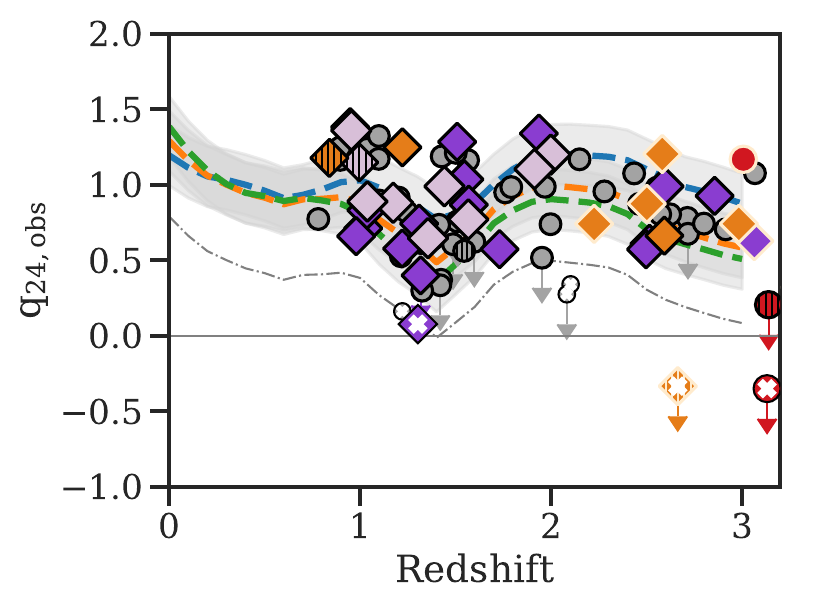}
\caption{The radio-infrared correlation for the full radio sample.  Symbols are as in Figure~\ref{fig:xray}. The dashed lines indicate representative SFG templates \citep{rie09, ruj13} at log $L_{\rm IR}/\Lsun$ = [11, 11.5, 12], with an assumed 0.2 dex scatter (shaded regions).  The dot-dash line marks our criterion for an outlier, 0.5 dex below the mid-point of the SFG locus as defined by the templates; a source below this line is classified as a radio-loud AGN. This criterion is applied up to $z=3$, where the 24$\,\mu$m band probes SF. The solid line indicates the canonical cut at q$_{24,\rm obs}=0$ adopted in previous studies \citep{don05}. 
}
\label{fig:q24}
\end{figure}

Figure~\ref{fig:q24} shows q$_{24,\rm obs}$ for our sample up to $z\sim3$, the highest redshift where 24$\,\mu$m traces SF activity. To define a SFG locus, we extract q$_{24,\rm obs}$  over $0<z<3$ for a set of local representative SFG templates from \citet{rie09} at log $L_{\rm IR}/\Lsun$ = [11, 11.5, 12]. These templates form a reasonable basis for infrared bright SFGs with a scatter of $\sim0.2$ dex in SFR \citep{rie09} up to $z\sim3$ \citep{ruj13,derossi18}.  We define an outlier from this distribution as being 0.5 dex below the mid-point of this locus (which is approximately represented by a log $L_{\rm IR}/\Lsun = 11.5$ template, Figure~\ref{fig:q24}). This method is adopted as recent works have shown that previous uses of $q_{24}$ in the literature such as adopting a threshold \citep[q$_{24,\rm obs}<0$;][]{don05}
or an archetypal template \citep[M82;][]{bon13} can result in significant incompleteness \citep{delm13, del17}.  Using this criterion, we identify three sources with radio-excess relative to the SFG q$_{24,\rm obs}$, one of which is also an X-ray identified AGN.

As noted in Section~\ref{sec:otherdata}, 11/100 of our sources are blended at the resolution of MIPS and so excluded from this analysis.  An additional 13 radio sources are undetected in MIPS.  We evaluate these sources for radio loudness by estimating their $3\sigma$ upper limits.  The GOODS-S 24$\,\mu$m map has a nominal rms of 4$\,\mu$Jy; however, at this depth, confusion noise will drive the achievable detection limit. As discussed in \citet{dole04}, confusion noise is a function of local source density and beamsize, raising detection limits at this depth by $2.5-3$ times the instrument noise.  For our data, an accurate source position was used as a prior, resulting in a slightly smaller confusion noise component.  As the local confusion limit is difficult to estimate quantitatively, we assume a factor of two times the rms level and show the $3\sigma_{\rm conf}$ upper limits in Figure~\ref{fig:q24}.  We find three MIPS-undetected radio sources which are indicated to be radio-loud AGN.  Two of these sources were already identified as AGN; of these, VLA033243.3-274646 was already known to be an outlier from the correlation, via ALMA observations \citep{ruj18}. The third, VLA033229.8-274400, is more tentative; it is a weak ($\sim4\sigma$) detection at 6 GHz with a marginally constraining MIPS upper limit indicating radio excess.  We include it in our sample but caution that confirmation is needed. The radio excess sources are listed in Table~\ref{tbl:simple}.

%\begin{figure*}[ht!]
%\includegraphics[width=0.45\textwidth, trim=2mm 0 0 0, clip]{figures/radio_ir_corr_fixed.eps}
%\includegraphics[width=0.45\textwidth, trim=2mm 0 0 0, clip]{figures/radio_ir_corr_actual.eps}
%\caption{\comment{(left) Fixed alpha. (right) Measured alpha. See text for description.}}
%\label{fig:radioir}
%\end{figure*}

\subsubsection{The Radio Spectral Slope}\label{sec:index}

In the previous two sections, we examined our radio sample for indications of radio-loud AGN activity.  The majority of AGN, however, are radio-quiet and have radio emission that originates from star formation and/or the AGN via, e.g., small-scale jets or outflows.  In both the RL and RQ populations, an AGN embedded in a compact, optically thick radio core will experience synchrotron self-absorption \citep{ryb86}, which will flatten the radio slope relative to that from star formation or an optically thin AGN.  This provides a relatively clean AGN selection, with radio slopes $\alpha\geq-0.5$, termed ``flat spectrum source" (FSS) AGN \citep[e.g. ][]{wal75, pea81, wil01, kim08, mas11, pad16, pad17}.

\begin{figure}[ht!]
\hspace{-3mm}
\includegraphics[scale=0.67, trim=2mm 0 2mm 0, clip]{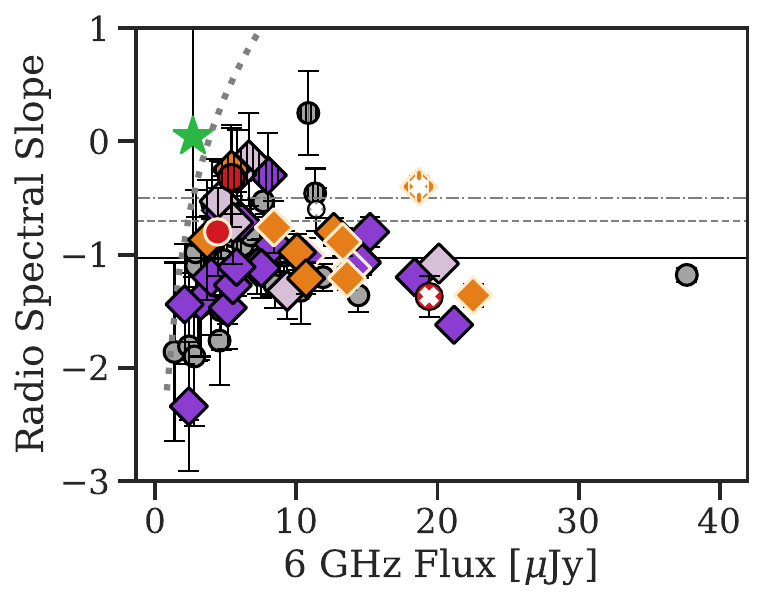}
\caption{The radio spectral slope of our sources as a function of their 6 GHz flux densities. Symbols are as in Figure~\ref{fig:xray}. The dotted line indicates our sensitivity limit for the radio slope based on the shallower 3 GHz data.  The green star is the stack of the 26 sources not detected at 3 GHz.  
The solid line indicates the average slope of our radio sample with detections at both 3 and 6 GHz.  The dashed line indicates the canonical $\alpha=-0.7$ slope for SFG \citep{con92}, while the dot-dashed line indicates the boundary for flat spectrum AGN $\alpha\geq-0.5$ \citep[i.e.][]{pad16,pad17}.}
\label{fig:slope_flux}
\end{figure}

In this section, we measure the radio spectral slope, $\alpha$, assuming a power law spectrum using our (observed) 3 and 6 GHz photometry for the 74 sources that have detections in both bands.  Although the beam size differs by up to a factor of $\sim$ 2 between our radio bands, \citet{gim19} show that we can expect this to have a minimal effect on our determination of the spectral slopes, further mitigated by our use of tapering. However, requiring a detection in both bands will bias us against flat radio slopes and observational limitations may result in artificially steepened slopes; we discuss these potential biases in the next section.  The majority of our sources are at $0.75<z<3$, which corresponds to a rest frequency range of 5-24 GHz. Figure~\ref{fig:slope_flux} shows our derived radio spectral slopes as a function of the 6 GHz flux.  We find that sources detected in both bands have $<\alpha>=-1.0\pm0.5$, an average that is consistent within the scatter with the canonical value of $\alpha=-0.7$ \citep{con92} and in good agreement with recent works that find a steep ($\alpha\sim-1$) slope at high frequencies \citep{tis19}.  
%We note, however, potential biases in this analysis: that observational limitations may result in artificially steepened slopes

\begin{figure*}[ht!]
\centering\includegraphics[scale=0.75, trim=2mm 0 0 0, clip]{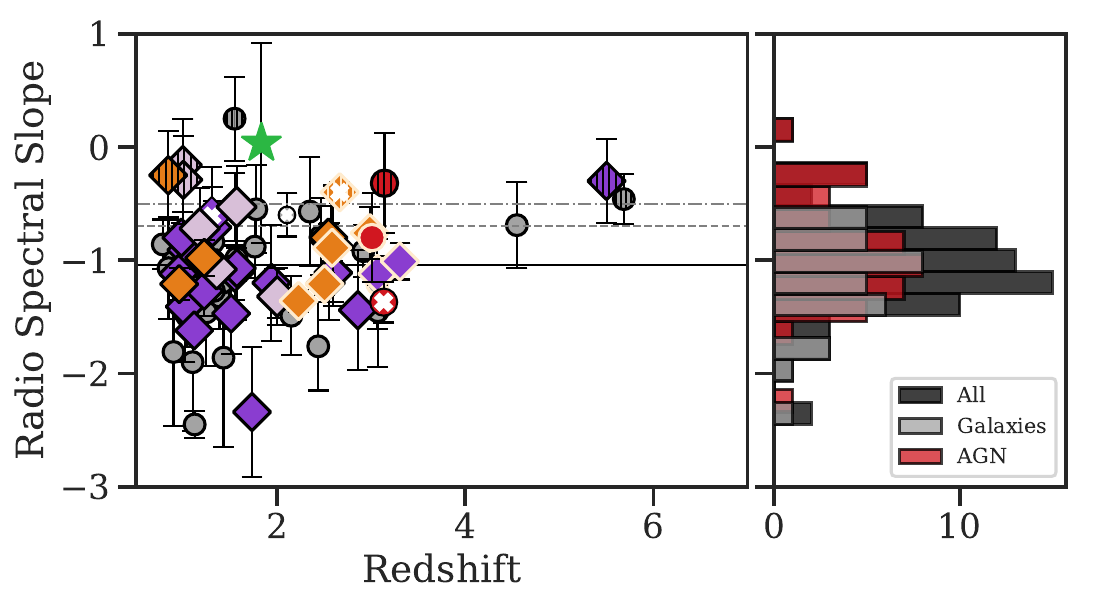}
\caption{{\it (left)} The radio spectral slope as a function of redshift. Symbols are as in Figure~\ref{fig:xray}.  The green star is the stack of the 26 sources not detected at 3 GHz.  Lines are as in Figure~\ref{fig:slope_flux}. {\it (right)} Histogram of all radio sources (black), those classified as galaxies (gray), and those classified as AGN (red).  There is no significant difference in the average slope of the galaxy and AGN 3 GHz-detected subsamples.
}
\label{fig:slope_z}
\end{figure*}

For the quarter of our sources not detected at 3 GHz, we measure an average radio slope by performing median stacking on the 3 GHz image using the publicly available code {\tt Stacker} \citep{lin15}.  We find that the 6 GHz sources undetected at 3 GHz have a stacked flux density of $S_{3 \rm GHz}=1.8\pm0.4 \,\mu$Jy.  The stacked error has been increased by $20\%$ due to stacking on the image rather than in the $uv$-plane, as suggested in \citet{lin15}. The resulting slope based on the stacked 3 GHz and the median 6 GHz photometry can be seen as the star in Figures~\ref{fig:slope_flux}, \ref{fig:slope_z}.  On average, our sources undetected at 3 GHz fall just below the detection limit. They may have shallower slopes than the rest of our sample, although the result differs from an on-average steep spectrum by only $\sim2\sigma$ and needs to be refined with deeper radio data at 3 GHz, which will be presented in future work.

Though low redshift studies over a comparable frequency range have recently found a significant fraction of FSS AGN within luminous Seyfert and LINER populations \citep{zaj19}, the expected fraction of FFS AGN is not yet well constrained, particularly in the {\it faint} RQ regime.  Our analysis reveals 8 FSS AGN candidates (Table~\ref{tbl:simple}), 6 of which were previously identified as AGN and one of which is RL. 
This number is likely a lower limit as the depth of the 3 GHz data biases our analysis against flat radio slopes (see Section~\ref{sec:bias}).  In Figure~\ref{fig:slope_z}, we look at the distribution of radio slopes as a function of redshift, finding no redshift dependence, and break the radio spectral slopes into the galaxy and AGN sub-populations.  We find that, minus the FSS AGN, the average slopes are identical, $<\alpha>=-1.1\pm0.4$.
%, suggesting that for the majority of our radio sample, the radio emission for non-AGN and AGN is powered by star formation. 
The source of the radio emission is discussed further in Section~\ref{sec:disc}.

Given that our observations are at high rest-frame frequencies, we consider whether our flat radio spectral indices could be produced by mechanisms unrelated to AGN activity.  At $\nu\gtrsim10$ GHz, it is expected that free-free emission plays an increasingly important role, adding a flat ($\alpha_{ff}=-0.1$) thermal component to the steeper synchrotron spectrum that dominates at lower frequencies \citep{kle88, con92, tab17}.  This effect would make the identification of AGN via a flat spectrum more ambiguous with our data.  In practice, however, a flattening of the radio spectral slope is not observed in high redshift SFGs.  As mentioned earlier, \citet{tis19} recently showed that, for a sample of star forming galaxies (SFR $>10\,\Msun$ yr$^{-1}$) up to $z\sim4$, the average rest-frame 0.5-15 GHz radio SED is best described by a broken power law with a steep ($\alpha\sim-1$) slope at high frequencies (rest $\nu>4.3$ GHz). A detailed analysis of 14 local star-forming galaxies by \citet{klein18} using radio data up through 24.5 GHz provides a closer look at this behavior. First, low-mass, low-luminosity galaxies do indeed have a large free-free component in their GHz radio spectra, but the fraction drops substantially for massive, luminous galaxies \citep[such those in our sample, see also][]{cle08, ler11}. Second, the best-fit non-thermal spectra for their massive galaxies  {\it all}  steepened toward high frequencies, and in a way that preserves the general power law slope despite the presence of a significant free-free component.  We are therefore confident that free-free emission is unlikely to mimic a flat spectrum AGN in our data.

\subsubsection{Observational Biases Toward Steeper Slopes} \label{sec:bias}

Observational limitations of our datasets may result in 1) the inability to measure a slope for radio sources with relatively flat radio spectra and/or 2) the artificial steepening of the slopes we do measure.   The first effect stems from the depth of the 3 GHz data, which is of order two times shallower than the 6 GHz survey.  Requiring a detection in both bands to measure the radio spectral index produces a bias against flatter radio spectra.  In the previous section, we demonstrated that the stacked 3 GHz flux density of 3 GHz undetected sources lies just below our detection limit on average, indicating that the fraction of FSS AGN identified in this study is likely a lower limit.  

Artificial steepening of the measured slopes may, on the other hand, result from the underestimation of the true flux density due to an inability to recover emission on scales that are poorly sampled by the current $uv$-coverage.  This effect will disproportionally affect our 6 GHz imaging, which has a factor of two higher resolution than that at 3 GHz.  Recent work by \citet{gim19} demonstrated that measuring the radio spectral slope based on radio imaging surveys with significantly different resolutions can greatly increase the scatter in the measured slopes.  While our resolution difference is in a range where they find only a minimal increase in scatter, we test for this effect by looking at our radio spectral slopes as a function of the ratio between the 6 GHz peak radio flux from the native resolution map and the total radio flux after tapering (see Section~\ref{sec:radiodata}).  For compact or point sources, this ratio will be one. If extended sources are preferentially having flux resolved out at 6 GHz, we should see steeper slopes for more extended sources.  From this test, we ascertain that there is no correlation between radio slope and this proxy for source compactness in our data.

While the above test gives us confidence in our measured photometry for extended sources, we cannot entirely rule out a frequency-dependent underestimation of the radio fluxes, as tapering cannot recover faint, diffuse emission in the case of extremely poor $uv$-coverage. A simple test was performed to look for a correlation between radio spectral slope and the SNR of the 6 GHz detections.  Steeper slopes at low SNR would potentially indicate losses due to poorly covered baselines; we find no such trend in our data.  

We highlight one additional secondary effect which may affect our data: in this work, we have utilized the multi-term multi-frequency synthesis (MT-MFS) algorithm to account for the frequency-dependent sky brightness of our broadband data and the $w$-projection algorithm to correct for widefield errors caused by non-coplanar baselines.  This is followed by a post-deconvolution wideband primary beam correction.  We note that this imaging strategy does not correct the second order flux and spectral index errors caused by the rotation of the primary beam pattern of the VLA antennae with time (the so-called $A$-terms).  These uncorrelated direction-dependent primary beam errors are known to cause an artificial steepening of source spectral indices by approximately $-0.1\pm0.2$ at the half-power point of the primary beam \citep{rau16}. A more advanced, but currently computationally prohibitive, strategy that would reduce the artificial spectral index steepening would be to employ the full $aw$-projection algorithm \citep{jag18}.   A more rigorous elimination of the potential spectral index limitations and biases discussed in this section will require additional observations with shorter baselines at 6 GHz and the application of a more advanced algorithm.  For this work, we stress that in spite of the caveats discussed in this section, our imaging strategy is sufficient to establish an interesting lower limit on the occurrence of FSS AGN signatures.   

\subsection{X-ray AGN Without Radio Detections}\label{sec:x_radio_undetect}

To assess fully the source density of known AGN in the ultra-deep VLA area, we  expand our search to include X-ray identified AGN in the \citet{luo17} catalog that are radio-undetected. For these X-ray AGN, in addition to having no radio counterpart, we impose the same criteria as in Section~\ref{sec:radiodata}, namely $z>0.75$ and an optical/NIR counterpart in the CANDELS/3D-HST catalog, identified by \citet{luo17} using their likelihood-based counterpart matching.  We find 57 radio-undetected X-ray AGN.

\section{Discussion} \label{sec:disc}

In the preceding sections, we have applied a multi-wavelength approach to identify AGN in a radio-selected sample from ultra-deep X-ray, optical-MIR, and radio imaging.  We identified AGN in fully half ($51\pm10\%$) of our sample of 100 radio sources.  Thanks to ultra-deep X-ray imaging with high completeness \citep{luo17}, the majority of these AGN (41/51 or $\sim80\%$) have some AGN signature in the X-ray: either an X-ray luminosity indicative of an AGN, a hard X-ray spectrum, and/or an excess in X-ray to optical, near-IR, or radio emission.  AGN are additionally identified through MIR excess, using MIR colors and optical-MIR SED fitting, and through radio properties.

Our parent sample is based on ultra-deep, high resolution radio imaging covering 42 arcmin$^{-2}$ and so consists of faint radio sources mainly in the $\mu$Jy regime, S$_{\rm 6 GHz}\sim 1-40\,\mu$Jy, corresponding to S$_{\rm 1.4 GHz}\sim 3-110\,\mu$Jy assuming a standard $\alpha=-0.7$ slope. It is the first opportunity to probe the radio properties of AGN at this faint level, reaching 4-10$\times$ deeper than comparable studies.  In the EDCF-S region, which encompasses the GOODS-S/HUDF region and our survey, deep VLA 1.4 GHz measurements \citep{sey04, mil13} enabled multiple studies of AGN in radio populations down to detection limits of S$_{\rm 1.4 GHz}\sim30-40\,\mu$Jy.  Utilizing infrared and radio properties to identify AGN, \citet{sey08} measured an AGN fraction of $28\pm8\%$.  Later work included X-ray and q$_{\rm 24,obs}$ indicators, finding an AGN fraction of $43\pm4\%$ \citep{bon13, pad15}.  A direct comparison requires a few caveats: our sample is restricted in redshift range ($z>0.75$) and covers a smaller area, biasing us against rarer, luminous sources.  Nevertheless our result is in good agreement with the latter studies at 1.4 GHz, demonstrating that the AGN fraction remains substantial at fainter radio fluxes.

We similarly compare to recent results from the VLA-COSMOS 3 GHz Large Project, which reaches a detection limit equivalent to S$_{\rm 1.4 GHz}\sim20\,\mu$Jy \citep{smo17, del17}.  AGN in this 3 GHz survey were subdivided into two categories: moderate-to-high radiative luminosity AGN (HLAGN) and low-to-moderate radiative luminosity AGN (MLAGN), where luminosity refers to the AGN luminosity and is a proxy for the radiative efficiency of BH accretion \citep{del17}.  The former are identified via X-ray, MIR, and/or optical-MIR SED fitting, while the latter are radio excess AGN that are not also HLAGN. Using the COSMOS 3 GHz sample, \citet{del17} identified $21\%$ of the sample as HLAGN and $17\%$ as MLAGN.  By contrast, we find $46\pm10\%$ of our AGN are HLAGN, and only $3\pm10\%$ MLAGN (Table~\ref{tbl:summary}).  This difference is likely primarily due to the completeness of X-ray data available in the GOODS-S/HUDF, highlighting the importance of ultra-deep X-ray in not only identifying but characterizing AGN. We can also compare the fraction of HLAGN with radio excess: $\sim30\%$ for the COSMOS sample and only $\sim7\%$ in our survey, possibly due to the smaller area covered and depth of our radio imaging.  If we expand the classifications presented above to include any radio signature (radio excess {\it and} flat radio spectra), then our numbers increase slightly to $5\%$ MLAGN and $17\%$ HLAGN with radio excess and/or a flat radio spectrum; however, it is unclear that these signatures should be grouped in terms of the represented mode of BH accretion \citep{whit17}.

In the following sections, we first discuss the majority of our AGN sample to address what is powering the radio emission in radio-quiet AGN.  We then turn the discussion to AGN that are indicated directly via radio indicators, i.e. radio loudness and a flat or inverted radio spectral slope. Finally, we discuss a number of items that pertain to the future possibilities for improving our understanding of the overall AGN population.

\subsection{The Origin of Radio Emission in High-z, Radio-Quiet AGN} \label{sec:hostgals}

The radio emission of faint RQ AGN at high redshift is thought to be dominated by star formation. Modeling suggests that SFGs dominate the total 1.4 GHz counts below $\sim100\,\mu$Jy \citep{man17}, or correspondingly below $L_{6 \rm GHz}\sim10^{23.5}\,\Lsun$ at $z\sim2.5$.  Radio stacking of X-ray identified AGN supports this conclusion \citep{pierce2011}.
Our study, however, is the first instance where this question can be probed with radio data sufficiently sensitive to {\it detect} typical SFGs at $1 \le z \le 3$ \citep{mad14}.
In the following sections we approach the question of the source of radio emission in AGN by:  {\bf (1)} investigating whether the radio flux densities are as expected for star forming galaxies; {\bf (2)} showing that typical radio-quiet AGN are not expected to contribute significantly to the radio at these flux densities; and {\bf (3)} showing that the galaxy morphologies are typical for SFGs.  From these analyses, we conclude that, indeed, the radio emission of AGN in our sample is dominated by the star formation in their host galaxies.

\subsubsection{Radio-Infrared Relation for Star-Forming Galaxies at 160$\,\mu$m} \label{sec:q160}

In addition to identifying outliers, the radio-infrared relation can be used to test whether radio flux densities are consistent with originating from star formation activity. In Section~\ref{sec:radio_ir}, we utilized 24$\,\mu$m measurements to look for outliers since this band matches the radio depth as closely as possible; here, we focus on measurements at observed 160$\,\mu$m, as this wavelength regime is dominated by dust heated by star formation over the entire redshift range of interest (1 $\lesssim z \lesssim 3$), with little to no contribution from AGN emission \citep[i.e.][]{kir13}.

Approximately half of our radio sample is detected at 160$\,\mu$m, a rate expected due to the incompleteness driven by confusion noise in the {\it Herschel} data at these flux levels \citep{ber13}. For this sub-sample, we first determine $q_{160}^{\rm 6 GHz}=$ log($S_{160,\rm obs}/S_{6 \rm GHz,obs}$).  For upper limits on the remaining sources, we adopt a detection limit of 5.2 mJy, which corresponds to the 80\% completeness for this field \citep{berta11}. We find our sample has a distribution with a standard deviation of 0.35 dex.

For comparison, we use a fiducial SFG: the log $L_{\rm IR}/\Lsun$ = 11.5 template of \citet{rie09}. This template has been shown to be representative of the SEDs of infrared galaxies at 1 $<$ z $<$ 3, the redshift range of our sample \citep{ruj13, derossi18}.  To check the appropriateness of our fiducial model, we compare the $q_{160}^{\rm 1.4 GHz}$ derived from the template to a SFG sample from \citet{mao11}, adjusting their $q_{70}$ using a color correction from our fiducual \citet{rie09} template.  We find that the log $L_{\rm IR}/\Lsun$ = 11.5 template well represents the \citet{mao11} sample. We therefore use the template to generate redshift-dependent values of $q_{\rm 160,fiducual}^{\rm 1.4 GHz}$, which we correct from 1.4 to 6 GHz assuming a radio spectrum of $\alpha=-0.8$ to obtain $q_{\rm 160,fiducual}^{\rm 6 GHz}$ ($=2.71, 2.67, 2.38$ at $z=1,2,3$, respectively)\footnote{The full redshift dependence of $q_{\rm 160,fiducial}^{\rm 6GHz}$ for $0.7<z<3.4$ can be reproduced using the following polynomial: $2.79 - 0.516\times z + 0.748 \times z^2 - 0.367\times z^3 + 0.0533\times z^4$.}.  
To determine the width of the distribution around $q_{\rm 160, fiducial}^{\rm 6 GHz}$, we evaluate the distributions for local, high metallicity galaxies for $q_{160}^{\rm 1.4 GHz}$ from \citet{qiu17} and for $q_{60}^{\rm 1.4 GHz}$ from  \citet{yun01}. At $z\sim 0$, these two wavelengths $-$ 160 and 60$\,\mu$m $-$ roughly bracket the rest wavelength range for observed 160$\,\mu$m for our sample. We find a virtually identical width to that for the high-redshift galaxies, with a standard deviation of 0.36 dex. 

\begin{figure*}[t]
\centering
\includegraphics[scale=0.75, trim=2mm 0 0 0, clip]{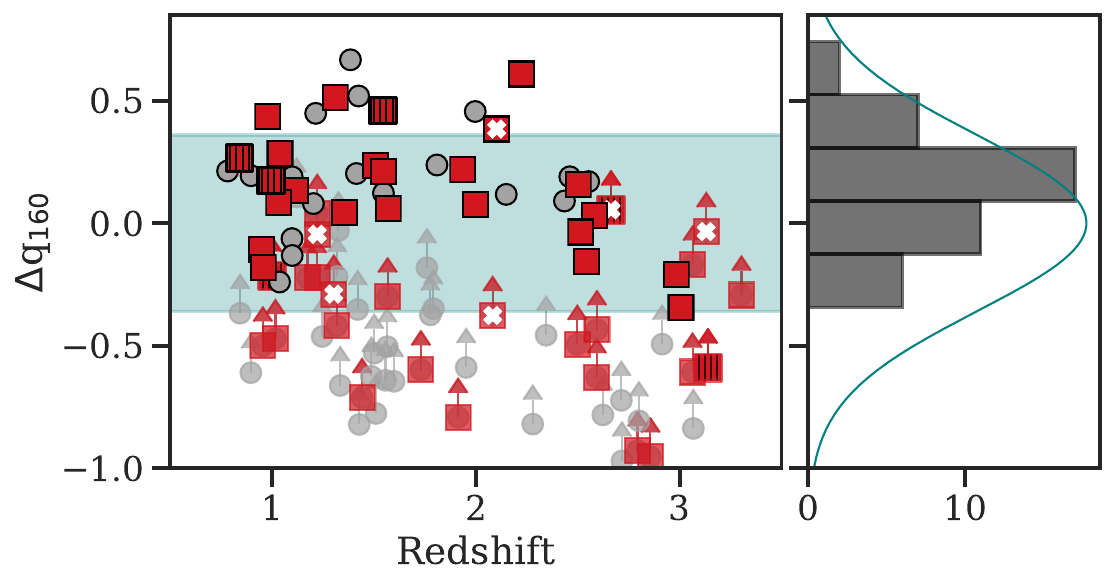}
\caption{{\it (Left)} The deviation, $\Delta q_{160}$, of the radio-infrared correlation at 160$\,\mu$m for our radio sources from a fiducial SFG template (Section~\ref{sec:q160}) as a function of redshift (three sources at $z>3.5$ are not shown for clarity).  The symbols show AGN (red squares) and galaxies (gray circles), with $3\sigma$ lower limits.  FSS and RL AGN are indicated as in Figure~\ref{fig:xray}.  The shaded region denotes the $1\sigma$ scatter found in local populations \citep{yun01,qiu17}. {\it (Right)} Histogram of  $\Delta q_{160}$ for all detected sources.  The overlaid Gaussian is normalized (not fit) to the maxmimum height of the histogram, centered on zero with the local scatter.
}
\label{fig:q160}
\end{figure*}

We show the distribution of values for the GOODS-S sample compared with the predicted values determined from the log $L_{\rm IR}/\Lsun$ = 11.5 template in Figure~\ref{fig:q160}, where $\Delta q_{160} = q^{\rm 6 GHz}_{\rm 160, fiducial} - q^{\rm 6 GHz}_{\rm 160, obs}$.  Sources that fall significantly   above the $\Delta q_{160}$ distribution would have excess radio relative to their 160$\,\mu$m emission.  The shaded region denotes $\pm$ the standard deviation found in local galaxies \citep{yun01,qiu17}, which is also displayed as a Gaussian distribution in comparison to the 160$\,\mu$m detected sources in the right panel.  The Gaussian has been normalized to the height of the source histogram, but not otherwise fit to the data.  It is nevertheless in good agreement, with a slight asymmetry which is likely due to the non-detections at 160$\,\mu$m. That is, the behavior of $\Delta q_{160}$ for our sample appears to be consistent with the scatter for radio emission powered purely by star formation.

 As discussed in Section~\ref{sec:radio_ir}, a typical criterion for a radio-loud AGN is that its ratio of infrared to radio should be a factor of ten or more lower than the typical value for star forming galaxies; we have shown that none of the 160$\,\mu$m detected sub-sample falls into this category.   However, we cannot rule out radio excess among the lower limits and, indeed, the majority of the RL AGN identified previously via their $q_{24}$ are found in this sub-sample.  However, our $q_{24}$ analysis shows that RL AGN are rare.
 Therefore, we conclude that, if the 160$\,\mu$m emission from these galaxies is powered as expected purely by star formation (and thereby 160$\,\mu$m non-detections are not biased toward or against radio excess from AGN), then most of the radio emission probably has the same origin. This behavior is consistent with that of brighter radio-quiet, or more descriptively non-jetted AGN \citep{pad16}, that often have far-infrared to radio ratios similar to those for star forming galaxies, leading to the belief that their radio outputs are dominated by star formation \citep[e.g.,][]{hec14}.

\subsubsection{Radio Output of Non-Jetted (Radio-Quiet)  Quasars}

It is still possible within the scatter of the SFG radio-infrared relation to find an AGN that contributes significantly to the total radio emission. However, we show in this section that the radio output from an AGN in our galaxies is likely to fall considerably below the output from star formation. For this purpose, we determine the radio-infrared relation for non-jetted, radio-quiet quasars (RQQs) in the sample of  \citet{white2017}, which they demonstrate have sufficiently luminous AGN to dominate ($>50\%$) in the radio. 

The sources in the \citet{white2017} sample have measurements at 1.4 GHz (VLA), 160$\,\mu$m ({\it Herschel}), and 24$\,\mu$m ({\it Spitzer}). In addition, all targets are at a single epoch (z $\sim$ 1), simplifying $k$-corrections. For this sample, the {\it Spitzer} 24$\,\mu$m observations probe rest 12$\,\mu$m, which can be a good measure of both the total AGN luminosity \citep{spin1989} and star formation activity \citep{ruj13}. Although we expect the 24$\,\mu$m outputs of these sources to be dominated by the output of the AGN, star formation can still make a  contribution.
We estimate this contribution as follows. We assume that the flux density at 160$\,\mu$m is entirely due to star formation and, using the \citet{rie09} log $L_{\rm IR}/\Lsun$ = 11.5 template,  determine that the ratio of flux density at 160 to 24$\,\mu$m should be $\sim$ 60 at  $z\sim1$  for purely stellar-powered outputs. With this estimate, we can remove the SF component from the flux density of the RQQs at 24$\,\mu$m by removing 1/60 of that observed at 160$\,\mu$m to get an estimate of the portion due {\it only} to the AGN. 

From the values with the star formation contribution removed, we compute $q_{\rm 24, AGN}^{\rm 1.4 GHz}$ = log (S$_{\rm 24\mu m, AGN}$/S$_{\rm 1.4 GHz})$ for the AGN component of each source, obtaining a lower limit for the cases not detected in the radio. We utilize all targets with 1.4 GHz measurements with SNRs $\ge$ 2 and adopt 2$\sigma$ upper limits for the rest. The average for the \citet{white2017} quasar sample (including lower limits that are $>$ 1.5)  is $q_{\rm 24 \mu m,AGN}^{\rm 1.4 GHz}$ = 1.7, with a median of 1.8. Three quasars from \citet{white2017} stand out as having low values of $q_{24}^{\rm 1.4 GHz}$ and may be radio intermediate. If we cut them from the sample, the average and median both become 1.8. These values can be compared with $q_{24}^{\rm 1.4 GHz}$ = 1.0 for purely star forming galaxies at z $\sim$ 1. That is, for non-jetted AGN the radio flux density at rest 3 GHz is 5 - 6 times weaker relative to the output at rest 12$\,\mu$m than is the case for purely star forming galaxies. To the extent that both AGN and SF radio spectra are optically thin synchrotron spectra with similar slopes, this conclusion will hold roughly  independent of radio frequency. Since the rest 12$\,\mu$m output of a star forming galaxy is correlated with its bolometric infrared luminosity, we can conclude that the radio output due to AGN activity in the majority of our ultra-faint radio sample is generally  significantly smaller than the radio emission due to star formation.

\subsubsection{Near-Infrared Morphologies} \label{sec:morph}

\begin{figure}[ht!]
\plotone{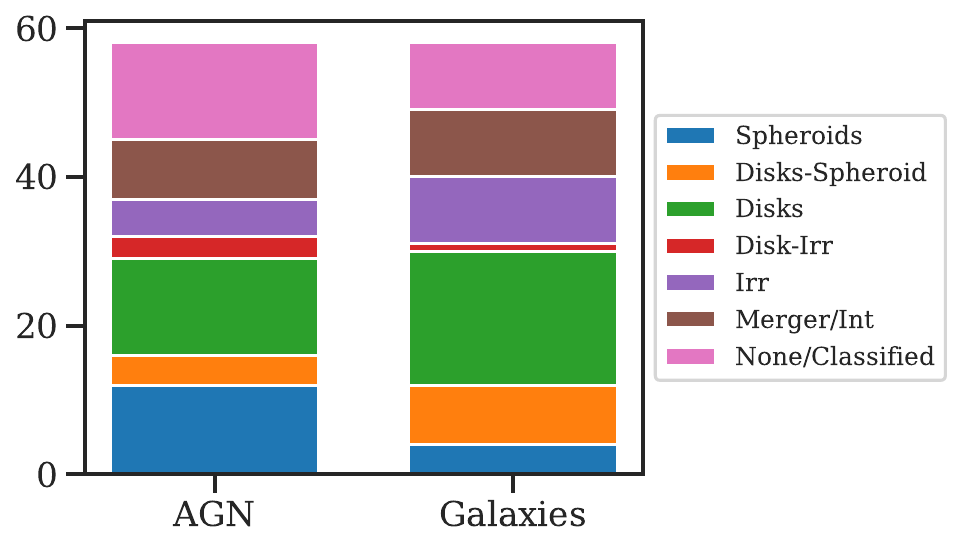}
\caption{The breakdown of the visually classified stellar morphologies \citep{kar15} for our AGN and galaxy sub-samples for the full faint radio population.  Stellar morphology classifications include spheroids, disks, and irregular (irr) galaxies, as well sub-classes combining those dominant classes.  None/unclassified includes sources where the source morphology could not be classified.  Sources are also flagged for potential merger or interaction activity; as this flag was in addition to the morphological classification, sources can be double counted in this representation.}
\label{fig:morph}
\end{figure}

To assess the morphologies of the host galaxies for our AGN sample, we adopt the visually classified near-infrared morphologies from \citet{kar15}.  These morphologies are based primarily on $H$-band imaging, with supplementary information from $V$- and $I$-band images, and are assessed by multiple classifiers.  The visual classification catalog includes multiple image depths; when possible, to be sensitive to disturbed morphologies, we adopt the classification in the deepest image.  Galaxies are classified into dominant types: spheroidal (including compact/point sources), disks, and peculiar/irregular, with combinations of these types possible, or none/unclassifiable.  Sources with signs of merger or interaction activity are flagged separately \citep[see][for more details]{kar15}.  Figure~\ref{fig:morph} shows the breakdown in our AGN and galaxy subsamples by visually classified morphology.  Overall, the morphology distribution is similar between the two subgroups, with AGN showing a slight preference for spheroidal hosts (by a factor of $3\pm1.7$).  Mergers and interactions make up a minority, $\sim15\%$, in both the AGN and galaxy subsamples, suggesting no excess in disturbed morphologies associated with AGN activity.  Keeping in mind sample size, we find no particular trend of morphology or interaction signature in our AGN sample, consistent with the hosts being comparable to the SFG population.
%with AGN selection technique.

\subsection{AGN Indicators in Faint Radio Sources: Radio-Loud and Flat Spectrum AGN}

In this work, we have identified RL AGN as outliers from the radio-IR correlation (Section~\ref{sec:radio_ir}) as well as both RL and RQ AGN as flat spectrum sources over the observed 3-6 GHz radio SED.  Within the overall AGN population, RL AGN are a minority \citep[$\lesssim10-20\%$;][]{bon13, wil15} and are typically associated with lower redshift, massive BHs at the centers of early-type galaxies.  Within our sample, we find 6/51 ($\sim12\%$) RL candidates with a range of redshifts ($z\sim1-3$), radio fluxes and spectral slopes, stellar masses (log $M_{\star}/\Msun\sim9.5-11$), and morphologies.  Half of our RL candidates are based on upper limits at 24$\,\mu$m, suggesting they may reside in hosts with lower SF, though we find that their visual morphologies are likely not early-type as is typical at lower redshifts.  The overlap of radio loudness with other AGN indicators is mixed: three are identified solely through radio excess, while the other three have X-ray and/or MIR signatures (see Table~\ref{tbl:simple}). One RL AGN is additionally a FSS AGN candidate.

The remainder of our AGN sample is radio-quiet, with 7 RQ AGN showing a direct indication of AGN activity through a flat radio spectrum.  Five of these FSS AGN are identified as AGN through multiple indicators, giving us confidence that they are bona fide AGN (Table~\ref{tbl:simple}).  The remaining two are tentative; even though SFGs at these redshifts show little evidence of free-free emission flattening out the radio spectrum at these frequencies \citep{tis19}, we cannot completely rule out non-obvious mechanisms which could mimic an AGN with a flat radio spectrum.  

We can compare the fraction of FSS AGN (8/51 or $\sim16\pm6\%$) in our faint radio sample to those from wider but shallower ATCA 5.5 GHz imaging in the GOOD-S field. There, \citet{huy15} found that $39\pm7\%$ of AGN with $50 \,\mu \mathrm{Jy} <S_{\rm 5.5 GHz}\,<100 \,\mu \mathrm{Jy}$ had a flat or inverted 1.4-5.5 GHz slope. 
Similarly, \citet{gim19} found a fraction of $30 \pm 8$\%  with slopes $\geq-0.5$, 
%using highly accurate slope measurements,
%\footnote{preferred over the GS one for the more accurate spectral indices}, 
in a radio sample largely fainter than 100$\mu$Jy at 5.5 GHz.  These fractions of FSS AGN are higher than ours by $\sim2.5\sigma$. This difference may in part be due to the quarter of our sample not detected at 3 GHz, which stacking suggests contains additional FSS AGN (Section~\ref{sec:index}).  Deeper 3 GHz imaging, to be presented in future work, will test this possibility.

Although it may be coincidental, the fractions found in faint radio populations \citep[][]{huy15, gim19} are similar to those for FSS AGN in much brighter samples, observed at similar frequencies, e.g. \citet{wal85}, \citet{zaj19}. These bright sources are generally associated with highly active AGN, e.g., blazars. The nature of the faint FSS AGN need not be analogous, but they are likely to be associated with an active nucleus \citep{gim19}. Extrapolations to low flux density radio AGN in empirically-based simulations suggest increasing core fractions with decreasing AGN luminosity, resulting in increasing numbers of FSS AGN due to, e.g., synchrotron self-absorption \citep{whit17}. This and the apparent uniformity of the fraction of FSS AGN suggests that flat radio spectra are useful indications of AGN activity across a broad range of radio flux densities. However, at faint levels the AGN nature of the candidates will need confirmation and the analysis must guard against  contamination by free-free emission.
 
In summary, out of 51 AGN in a radio-selected sample, about one-third have a direct indication of AGN activity in their radio properties.  This reiterates the need for a multi-wavelength approach in identifying AGN with current datasets and is consistent with radio emission in the $\mu$Jy regime being mostly associated with star formation in the host, which was addressed in detail in the previous section.

\subsection{The Source Density of AGN at Cosmic Noon} \label{sec:sourcedensity}

While sub-mJy radio-selected samples provide additional constraining information in a regime probing both SFG and AGN populations, the requirement of a radio detection introduces a bias against RQ AGN in more quiescent hosts. This bias is confirmed by the 57 additional AGN candidates found via their X-ray properties \citep{luo17} that are not detected in our radio data (Section~\ref{sec:x_radio_undetect}).

\begin{figure}[ht!]
\plotone{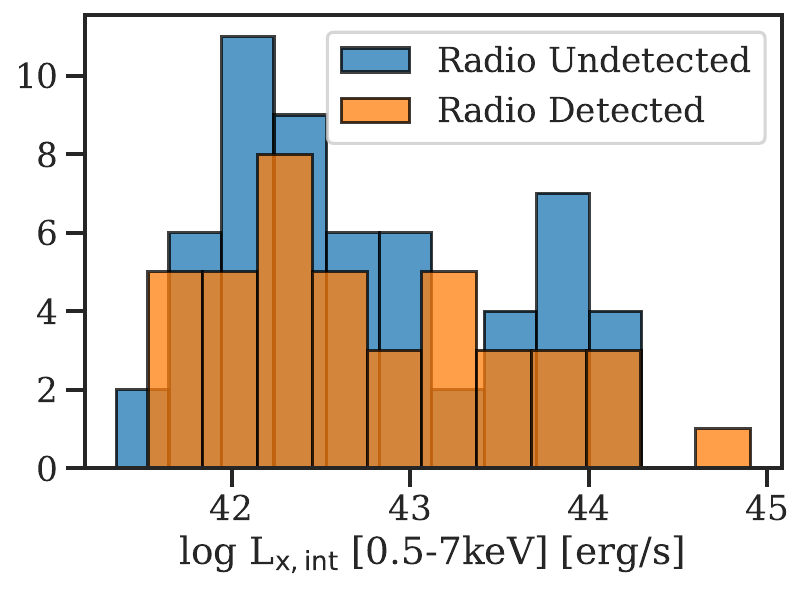}
\caption{Histogram showing the intrinsic X-ray luminosity for both radio detected (orange) and radio undetected (blue) AGN samples. }
\label{fig:xhist}
\end{figure}

The distributions of the intrinsic X-ray luminosities for our radio-detected and radio-undetected AGN samples can be seen in Figure~\ref{fig:xhist}; the distributions are qualitatively similar. 
Together the radio-selected sample of AGN and the radio-undetected X-ray AGN imply a source density of at least 2.6 arcmin$^{-2}$ at cosmic noon above $S_{\rm 0.5-7 keV}\sim 4\e{-17}$ erg s$^{-1}$ cm$^{-2}$.   
This does not include potential MIR excess AGN within the radio undetected population; however, as demonstrated in Section~\ref{sec:sedfitting}, the majority of AGN identified  ($\sim80\%$) have an AGN signature in the X-ray, due to the ultradeep X-ray available in the HUDF.  An additional missing population, heavily obscured AGN, will be discussed in the following sections.

\subsection{A Census of AGN: have we found them all?} \label{sec:census}

The primary goal of this work is to compile as complete a census of AGN activity as possible in an area of ultra-deep legacy surveys, using a multi-wavelength approach, with results 
summarized in Tables~\ref{tbl:summary}-\ref{tbl:simple}, Section~\ref{sec:disc}, and Section~\ref{sec:sourcedensity}.  In this section, we assess a particularly elusive AGN population that is almost certainly still underrepresented: heavily obscured AGN. We describe the nature of these objects, discuss future possibilities for their identification, and estimate the possible yields of such sources.

\subsubsection{Obscured AGN in this Work}

Obtaining a complete sample of AGN has been known to be a significant challenge for years.  Every search method finds only a fraction, as is made clear, for example, by the Venn diagram in \citet{del17}, the lack of edge-on galaxies in optically-selected samples \citep{mai95}, the failure of infrared methods to find all X-ray bright AGN \citep{don08}, and the failure of some infrared-identified AGN to be detected even with very deep X-ray data \citep[][this work]{del16}.  
In our study,
%we have found a small number of AGN through SED fitting that reveals an excess in the 3-5$\mu$m  range; 
the fraction of AGN found {\it only} in the MIR is small relative to the X-ray identifications, compared with the results of previous studies \citep[e.g.][]{don08, del17}. This difference is possibly because of the extremely deep X-ray data available in the GOODS-S/HUDF field.
However, the use of MIR SED fitting to identify AGN %routinely missed in X-ray surveys 
is {\it also} currently limited not only by the available photometric coverage, but by our understanding of the full range of diversity in intrinsic AGN spectra.  As discussed in Section~\ref{sec:sedfitting} and Appendix~\ref{app:modeling}, we limit our SED fitting primarily to a single Type-1 AGN template. Because of the broad range of SED predictions for Type-2 circumnuclear tori, we have not been able to include them in the fitting.

\begin{figure}[ht!]
\includegraphics[scale=0.67, trim=2mm 0 4mm 0, clip]{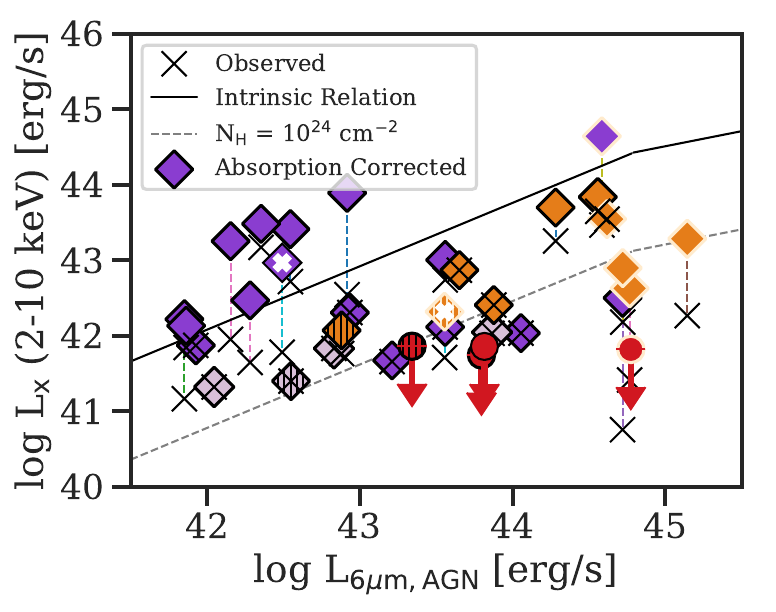}
\caption{The rest-frame 2-10 keV X-ray luminosity as a function of the rest 6$\,\mu$m luminosity of the AGN for our AGN that have sufficient photometry for SED fitting.  The full symbols are as in Figure~\ref{fig:xray} for the absorption corrected, intrinsic X-ray luminosity \citep[see][]{luo17}, which are connected by dashed lines to x symbols showing the uncorrected, observed X-ray luminosity.  The solid line indicates the intrinsic (unobscured) relation from \citet{che17}.  The dashed line is this relation obscured by a column density of N$_{\rm H}=10^{24}$ cm$^{-1}$; AGN below this line are in the Compton-thick regime.}
\label{fig:delmoro}
\end{figure}

This omission produces a bias prohibiting a complete census of obscured AGN.  Alternative techniques can somewhat make up for this bias; obscured AGN are revealed in our sample as 11 AGN with hard X-ray spectra, the 9 MIR PL AGN, and the handful of optical spectroscopic confirmations of high ionization narrow emission lines, which indicate a Type-2.  To  assess more accurately the fraction of moderately and heavily obscured AGN in our sample, we utilize the well-established correlation between the intrinsic X-ray luminosity and the near-infrared emission of the AGN \citep[i.e.][]{fio09, gan09, del16, che17}, here determined at rest 6$\,\mu$m from the best-fit AGN template (Figure~\ref{fig:delmoro}). To compare to the relations established in the literature, we derive the rest 2-10 keV X-ray luminosity using the {\tt Sherpa} python package \citep{fre01}.  We then compare to the intrinsic relation derived in \citet{che17}.  We additionally compare to this relation given a factor of 20 in attenuation of the X-ray, corresponding to a column density of N$_{\rm H}=10^{24}$ cm$^{-1}$ erg s$^{-1}$ \citep{lan15, lan17}; this column density marks the Compton-thick (CT) regime.

We find that our AGN largely scatter around the intrinsic relation, as expected, but with signs of increasing obscuration at higher 6$\,\mu$m luminosities.  About $30\%$ of our AGN are consistent with being heavily obscured or Compton-thick, which is within the range of predicted CT fractions \citep[$\sim10-50\%;$ i.e.][]{tre09, aky12}.  This range is, however, poorly constrained, as estimates of the CT fraction currently rely on X-ray observations, which suffer from degeneracies between obscuration and the X-ray reflection component \citep{akylas2016, geo2019}.

\subsubsection{The Bimodal Obscured AGN Population} \label{sec:obscured}

The most elusive AGN are those with Compton-thick absorbing columns, such as those found in the SWIFT-BAT survey by \citet{akylas2016}. To understand the prospects for finding a more complete sample of Compton-thick AGN at higher redshifts, we need to understand their nature. The BAT sample seems to include two types. The first is cases where we view the AGN almost exactly edge-on and the high column is because 
we are looking edge-wise through the circumnuclear torus. A good example is NGC 5728, studied in detail by \citet{durre2018}. Their measurements show two prominent ionization cones in [OIII], 
very symmetric in appearance. At their apices is the core of the reflected X-ray emission that is also distributed symetrically, as well as the peak of the radio emission. The second case is where the 
AGN is deeply embedded and its energy is largely absorbed and re-radiated in the infrared. In extreme absorbed cases, the energy of the AGN is completely thermalized 
and emerges as a $\sim$ 300 K blackbody-like spectrum, albeit with the effects of radiative transfer, such as strong silicate absorption. Mrk 231 and IRAS 08572+3915 are two possible examples \citep{armus2007}. It is beyond the scope of this paper to test the prevalence of these two types in the BAT sample, but it appears that both types are well-represented. 

However, one must be careful not to over-simplify any classification scheme. As an illustration, Mrk 231 is apparently not Compton-thick from the {\it NuSTAR} spectrum, with $N_{\rm H} = 
1.2 \times 10^{23}$ cm$^{-2}$ \citep{teng2014}. However, its nuclear SED in the infrared strongly suggests a pseudo-blackbody,  as expected if 
nearly all of its energy has been absorbed, and its complete lack of mid-infrared emission lines \citep{armus2007} is consistent with this hypothesis. \citet{teng2014} suggest that the lack of lines is due to an intrinsically weak UV and X-ray continuum, but the lack of {\it all} lines, including for example all of the forbidden neon lines with ionization potentials ranging from 21.6 eV (Ne II) to 126 eV (Ne VI), and of other bright mid-IR fine structure lines having even lower values (e.g., 15.8 eV for [Ar II], only slightly more than for hydrogen, 13.6 eV),  makes this explanation unlikely.  It seems more plausible that the circumnuclear material is highly clumpy and that our line of sight is relatively unobscured but not representative of the average obscuration over 4 $\pi$ sterradians, i.e., the source could be Compton-thick viewed from most directions but not in ours.

In any case, the example of Mrk 231 shows that the infrared behavior may not track exactly that in the X-ray, i.e., sources classified as Compton-thin (albeit with a substantial atomic column) from the X-ray spectrum might have the great majority of their luminosity absorbed by circumnuclear material, so that it emerges in the mid-infrared. Such sources, heavily absorbed and underluminous in the X-ray \citep{teng2014} would be under-represented in X-ray-selected samples, as would bonafide Compton-thick sources. In the following discussion we will describe the two types of behavior as ``absorbed" and ``edge-on" to avoid any reference to specific X-ray properties. 

We first consider the absorbed case. The bolometric luminosity of Type 1 AGN is an order of magnitude greater than their 0.5 - 10 keV luminosity \citep{lusso2012}. That is, an AGN with an X-ray luminosity $\ge 3 \times 10^{42}$ ergs s$^{-1}$ should have a bolometric luminosity $\ge 10^{10}$ $L_\odot$. AGN templates $-$ both those representing Type 1 \citep[e.g.][]{fritz2006} and Type 2 \citep[e.g.][]{nenkova2008}, the latter of which have been shown to fit partially obscured AGN relatively well \citep[see][]{gon19b} $-$ indicate that the SED of these cases is broad but generally peaked near 10$\,\mu$m, as demonstrated in the examples of Mrk 231 and IRAS 08572+3915.
%Templates,  \citep[e.g.][]{fritz2006, nenkova2008}, the latter shown to fit Type 2, i.e. partially obscured, SEDs relatively well \citep{gon19b}, indicate that the SED of these cases is broad but generally peaked near 10$\,\mu$m, as shown in the examples of Mrk 231 and IRAS 08572+3915. 
That is, the energy will emerge primarily in the mid-infrared and the SED will extend with considerable strength into the $3 - 5\,\mu$m range.  

To make this point quantitative, we take Mrk 231 as the prototype for an absorbed AGN. Adopting the relation between bulge and black hole mass from 
\citet{kor13}, equation 2, and taking the K - W2 color of a galaxy bulge to be 0.24 (from \citet{willner1984} plus our own calculation of W1 - W2 for a large sample of elliptical galaxies), we derive

\begin{equation}
\textrm{log} (L_{\rm Edd}/\Lsun) = 13.239 - 0.484(M_{W2} + 24.45)
\label{eqno01}
\end{equation}

\noindent
where $L_{\rm Edd}$ is the Eddington luminosity of an average mass black hole (according to the Magorrian Relation) in a galaxy bulge of absolute magnitude $M_{W2}$. The distance modulus of Mrk 231 is 36.38 and WISE measures W2 = 6.301, so its $M_{W2}$ = -30.1. From \citet{cutri1984} and the standard K - W2 color, the {\it stellar} $M_{W2}$ = -25.9, i.e., it can be ignored relative to the total at this band. The AGN luminosity is log $L_{\rm AGN}/\Lsun = 12.05$ \citep{farrah2003}, whereas equation~\ref{eqno01} gives log $L_{\rm Edd}/\Lsun = 13.94$. That is, the luminosity of the AGN is slightly more than 1\% of the Eddington luminosity corresponding to an average supermassive black hole in a galaxy of the stellar brightness of the Mrk 231 host. The AGN is, however, about 4.2 magnitudes brighter at 4.5$\,\mu$m than is the host galaxy, i.e., black holes emitting at $\ge$ 0.03\% of $L_{\rm Edd}$ with this luminosity absorbed by circumnuclear material should be identifiable through the excess emission at 4.5$\,\mu$m. This result should hold down into the Seyfert galaxy range, i.e., to AGN luminosities of $\sim 10^{10}$ $L_\odot$. 

We now consider the edge-on situation. There is a broad extinction minimum between 3 and 8$\,\mu$m \citep{xue2016}, where the nominal optical depths are similar to those at 6 - 8 keV in the X-ray \citep{corrales2016}. Reverberation mapping demonstrates that the infrared emission in this range originates at the inner rim of the circumnuclear torus, which is $\sim$ 0.1$-$2 pc in diameter depending on the AGN luminosity \citep{lyu2019}, whereas the X-ray source is expected to be extremely compact. Therefore, for a clumpy circumnuclear torus the 3$-$8$\,\mu$m infrared emission can escape along the relatively-less-obscured lines of sight over an extended region, and hence can escape more readily than is typical for the 6$-$8 keV X-ray emisison, given equal nominal optical depths. It is just in the 6$-$8 keV range that Compton-thick AGN are revealed \citep{geo2019}. This situation underlies one of the arguments that it should be possible to find ``hidden" AGN in the mid-infrared.  

To make a quantitative estimate, we take the ``normal'' quasar template of \citet{lyu2017}. If there is no extinction in the circumnuclear torus and the emission of its inner rim at 4.5$\,\mu$m is isotropic, the threshold for detection of the emission as an excess above the galaxy stellar spectrum is of order 0.1\% of $L_{Edd}$, again assuming the average supermassive black hole via the Magorrian Relation and using the measurements of Mrk 231 to set a conservative threshold. The zero extinction assumption is probably optimistic, but the assumption of isotropic emission is probably roughly correct. Assuming a net extinction of a factor of ten, the threshold is $\sim$ 1\% of $L_{Edd}$. 

\subsection{Future Detection Possibilities}\label{sec:future}

The Mid-Infrared Instrument on {\it JWST} will be very effective at finding absorbed AGN and also useful for edge-on ones, as illustrated by the estimates above. The multiple spectral bands in MIRI are well suited to probing galaxy SEDs for AGN activity up to $z\sim2.5$ \citep{mes12, kir17}, with sufficient spectral resolution to isolate the 4 - 5.5$\,\mu$m region that is a minimum in star forming galaxies, even those with prominent aromatic features.  A pilot program will be conducted by the US MIRI Guaranteed Time Observation (GTO) team (PI: G. Rieke) covering about 30 square arcmin in the GOODS-S/HUDF field overlapping with the deep VLA field discussed in this paper. The integration times have been set to obtain roughly equal signal to noise in the MIRI bands for a $\nu^{-2}$ power law between 5.6 and 21$\,\mu$m (excluding the filter at 11.3$\,\mu$m). At rest 6$\,\mu$m, this pilot survey will reach $5\sigma$ depths of 2$\,\mu$Jy (6$\,\mu$Jy) at $z\sim1$ ($z\sim2$) in the F1280W (F1800W) filter. Via the $L_x - L_{\rm 6\mu m}$ relation \citep[i.e.][]{che17}, this corresponds to an X-ray flux limit of $S_{\rm 0.5-7 keV}\sim5\e{-16}$ erg s$^{-1}$ cm$^2$ (log $L_{\rm 0.5-7 keV}/\mathrm{erg\,s^{-1}}\sim 42.3$) at $z\sim1$, assuming an unabsorbed X-ray source with $\Gamma=1.4$.  As an example, we should achieve signal to noise ratios in all these bands of $\gtrsim 10 : 1$ for sources with spectra similar to Mkn231 but only 1\% as bright at z = 1 (i.e., bolometric luminosity of $\sim 10^{10}$ $L_\odot$) and only 10\% as bright at z = 2 (i.e., bolometric luminosity of $\sim 10^{11}$ $L_\odot$).  

From the source density found in this study, we therefore expect to detect a minimum of $\sim50$ AGN in this survey at cosmic noon.  Given the ability to detect AGN down to $1\%$ of Eddington as previously discussed, the counts for obscured and/or low luminosity AGN may increase this number: recent work using infrared color diagnostics suggests a larger fraction of AGN among the IR-population detectable by MIRI, driven mainly by AGN-host composites with strong host emission in the low Eddington ratio regime \citep{kir17}.  Our MIRI GTO survey will test these predictions.  Taking our conservative estimate and the possible range of the fraction of Compton-thick AGN of 12 - 30\% of the total number of AGN \citep{akylas2016}, and assuming that half are absorbed, we then expect to find at least 3 - 8 examples in this survey, plus additional detections of edge-on cases.  Of course, it is possible that more (or less) examples will be found, depending on the properties of the AGN population. Future MIRI surveys of larger areas are important to improve the statistics regarding this population.

Are there other detection approaches? The signal to noise in the {\it NuSTAR} data \citep[e.g.,][]{geo2019} on the relatively nearby BAT sample indicates that using hard X-rays to characterize the distant population of Compton-thick and related AGN is beyond current technical capabilities. However, there are two additional possibilities. The first is to look for variability in the 3 - 5$\,\mu$m region; rms fluctuations of $\sim7$\% seem to be characteristic \citep{lyu2019}. The second is to search for galaxies with ratios of [OIII]$\lambda$5007 to H$\beta$ strongly indicative of an AGN but with no other indications. Such cases are candidates for edge-on systems with large columns of absorbing material from their circumnuclear disks.

\section{Conclusions} \label{sec:con}

This work presents an analysis of AGN in faint radio populations based on new, deep radio imaging at 6 (0.32 $\,\mu$Jy rms, 0.31$^{\prime\prime} \times $ 0.61$^{\prime\prime}$) and 3 GHz (0.75$\,\mu$Jy rms, 0.6\arcsec $\times$ 1.2\arcsec) in the GOODS-S/HUDF region.  Combining these with existing deep legacy datasets, we apply AGN identification techniques in the X-ray, optical-MIR, and radio to assess the AGN fraction and nature of AGN in a faint ($S_{\rm 1.4 GHz}\sim3-100\,\mu$Jy equivalent) radio sample.  Our main conclusions are as follows:

\begin{enumerate}

\item Within our 100 radio sources at $z>0.75$, we find that 51 are AGN candidates via one or multiple AGN indicators.  The majority ($80\%$) of our AGN are indicated via X-ray properties; the uniquely deep 7 Ms X-ray available in this field provides high completeness at the X-ray luminosities expected of moderately luminous AGN ($L_x\sim10^{42}$ erg s$^{-1}$).  15/51 have MIR excess, determined via MIR colors and/or SED fitting, including 5 candidates for warm-excess AGN, which may indicate a polar dust component \citep[i.e.][]{lyu18}.  6/51 have radio excess, indicating RL AGN activity, and at least 8/51 have flat or inverted radio spectral slopes over 3-6 GHz, indicating compact radio cores.  Additional flat spectrum source AGN are likely among the 3 GHz-undetected subset as indicated by stacking; it is plausible that flat radio spectra are as prevalent in our faint sample as found in brighter radio populations, making this a useful AGN indicator.  Deeper 3 GHz imaging to test this possibility will be presented in future work. 

\item In sources where radio emission is predominately from star formation activity, the X-ray to radio ratio provides a robust method for AGN identification.  Even given a conservative criterion of 5 times the maximal X-ray emission for a starburst (Section~\ref{sec:xrayradio}), combined deep X-ray and radio data yield 38/51 AGN identified via this method, 10 of which are not otherwise indicated by more traditional X-ray criteria, i.e. luminosity or X-ray hardness.

\item In good agreement with comparable studies \citep[4-10$\times$ shallower than our survey;][]{bon13, del17, cer18}, we determine that the radio output of our very faint RQ AGN are largely consistent with originating from star formation activity.  Comparing the radio-infrared ratio at 160$\,\mu$m of our AGN and SFG sub-samples to a fiducial SFG template \citep{rie09} recovers the local radio-infrared correlation for SFGs including the measured scatter from local samples.  This result is supported by the similar distributions of the radio spectral index between AGN and SFGs (minus FSS AGN) and by the comparable host morphologies and interaction rates between the two sub-samples.  We further demonstrate via a sample of luminous RQ (or non-jetted) quasars \citep{white2017} that for truly radio-quiet AGN, the output in the radio due to AGN activity is likely 5-6 times lower relative to the mid-infrared than that expected from star formation.

\item In addition to the 51 AGN in the radio-selected sample, 57 radio-undetected AGN are indicated via X-ray properties in the relevant area and redshift range.  Combined, these AGN populations indicate an AGN source density of at least 2.6 arcmin$^{-2}$ at cosmic noon above $S_{\rm 0.5-7 keV}\sim4\e{-17}$ erg s$^{-1}$ cm$^{-2}$.  This result is a lower limit given biases in our analysis against FSS AGN and heavily obscured or Compton-thick AGN. 

\item  Despite the success of multi-wavelength identification of AGN, a potentially substantial population of AGN, i.e. Compton-thick AGN, remains elusive.  Within our radio sample, we find that $\sim30\%$ are likely heavily obscured despite several biases against these sources in our selection techniques, i.e. relatively soft band X-ray observations and the assumption of a Type-1 AGN template during SED fitting. We discuss the nature of the potentially still missing Compton-thick AGN population, considering two likely cases of ``absorbed" and ``edge-on" configurations.  We show that in both cases, coverage of the rest 4-5.5$\,\mu$m region will provide a clear signature of AGN activity above the ubiquitous stellar minimum in inactive galaxies, down to $\sim$ 0.03$\%$ and $\sim$ 1$\%$ of $L_{\rm Edd}$ in ``absorbed" and ``edge-on" cases, respectively.  Coverage of these wavelengths will be provided by {\it JWST}/MIRI up to $z\sim2.5$.
\end{enumerate}

%% If you wish to include an acknowledgments section in your paper,
%% separate it off from the body of the text using the \acknowledgments
%% command.
\acknowledgments

The authors thank Jianwei Lyu and Benjamin Johnson for insights into SED fitting and Pablo P\'{e}rez-Gonz\'{a}lez for discussions on the {\tt Rainbow} Database.  The National Radio Astronomy Observatory is a facility of the National Science Foundation operated under cooperative agreement by Associated Universities, Inc. Basic research in radio astronomy at the U.S. Naval Research Laboratory is supported by 6.1 Base Funding. This research has made use of data obtained from the {\it Chandra} Data Archive and the {\it Chandra} Source Catalog, and software provided by the {\it Chandra} X-ray Center (CXC) in the application packages CIAO, ChIPS, and Sherpa. This work is based on observations taken by the 3D-HST Treasury Program ({\it HST}-GO-12177 and {\it HST}-GO-12328) with the NASA/ESA {\it Hubble} {\it Space} {\it Telescope}, which is operated by the Association of Universities for Research in Astronomy, Inc., under NASA contract NAS5-26555.  This work has further made use of the {\tt Rainbow} Cosmological Surveys Database, which is operated by the Centro de Astrobiología (CAB/INTA), partnered with the University of California Observatories at Santa Cruz (UCO/Lick,UCSC).  Additionally, this work has made use of {\tt ASTROPY}, a community developed core Python package for astronomy \citep{ast13} hosted at \url{http://www.astropy.org/}.

%% To help institutions obtain information on the effectiveness of their 
%% telescopes the AAS Journals has created a group of keywords for telescope 
%% facilities.
%
%% Following the acknowledgments section, use the following syntax and the
%% \facility{} or \facilities{} macros to list the keywords of facilities used 
%% in the research for the paper.  Each keyword is check against the master 
%% list during copy editing.  Individual instruments can be provided in 
%% parentheses, after the keyword, but they are not verified.

%\vspace{5mm}
%\facilities{HST(STIS), Swift(XRT and UVOT), AAVSO, CTIO:1.3m,
%CTIO:1.5m,CXO}

%% Similar to \facility{}, there is the optional \software command to allow 
%% authors a place to specify which programs were used during the creation of 
%% the manusscript. Authors should list each code and include either a
%% citation or url to the code inside ()s when available.

%\software{astropy \citep{2013A&A...558A..33A},  
%          Cloudy \citep{2013RMxAA..49..137F}, 
%          SExtractor \citep{1996A&AS..117..393B}
%          }

%% Appendix material should be preceded with a single \appendix command.
%% There should be a \section command for each appendix. Mark appendix
%% subsections with the same markup you use in the main body of the paper.

%% Each Appendix (indicated with \section) will be lettered A, B, C, etc.
%% The equation counter will reset when it encounters the \appendix
%% command and will number appendix equations (A1), (A2), etc. The
%% Figure and Table counter will not reset.

\appendix

\section{SED Fitting Details} \label{app:modeling}

SED fitting is performed using a non-negative linear combination of a small set of representative templates \citep{assef08, assef10, chu14, alb16}.  The templates include an elliptical model, representing the old stellar population, star forming models for young stellar populations, and an optional AGN component.  During fitting, a prior is applied to the $R$-band luminosity, empirically determined via data from the Las Campanas Redshift Survey \citep{lin96}, to avoid unphysical fits \citep{assef10}.  The IGM absorption is fixed as a function of redshift \citep{assef10,chu14}.  Additional reddening of the AGN template is allowed as a free parameter.

The \citet{assef10} SED fitting code comes pre-packaged with empirically-derived galaxy and AGN templates. However, likely because our galaxies were selected in a different manner than the original \citet{assef10} sample, we find these galaxy+AGN templates do not provide robust fits.  We make the following substitutions.  For galaxy templates, we replace the ``spiral'' and ``irregular'' templates with a a series of 16 spectra with variable optical attenuation (A$_v=0-3$) generated using the Flexible Stellar Population Synthesis code \citep[FSPS; ][]{con09, con10}. For more details on these spectra, see Appendix~\ref{app:fsps}.  These star forming spectra are then combined with the \citet{assef10} ``elliptical'' template, to allow for a variable old stellar population contribution relative to emission from young stars.  The best-fit of a single FSPS spectra + elliptical template is then adopted as the host galaxy fit. Figure~\ref{fig:fitex} shows an example.

After fitting the stellar-based  templates, we refit all sources with an added AGN template.  Here we adopt the empirical X-ray$-$radio AGN template presented in \citet{elv94}, modified to remove the host galaxy FIR contribution \citep{xu15}. The \citet{elv94} template supplies the intrinsic optical-MIR SED of a luminous Type-1 AGN \citep[see also][]{ric06, elv12, sco14, lyu16, lyu17, lyu17b}, which has been shown to be applicable to AGN and quasars over a range of redshifts \citep{jia06, jia10, xu15}, and properties such as Eddington ratio \citep{hao11, hao14}.  Reddening of the AGN template is allowed to vary as a free parameter. This approach does not include Type-2 AGN, which are expected to be just as prevalent.  Unfortunately, the SEDs of Type-2 AGN have yet to be robustly constrained, with Type-2 circumnuclear tori models producing a wide range of potential SEDs with strong degeneracies \citep[see][]{ram17, gon19a, gon19b}.  The photometric constraints in this work are not adequate to address these degeneracies and so Type-2 template(s) are not included in the SED models.

As discussed in \citet{chu14}, anomalous datapoints within large datasets can strongly bias fits. To test for this effect, we systematically remove each photometric band from our fittings, finding that no one datapoint is driving our fits.  However, we do find that to approach a median $\chi_{\nu}^2\sim1$, we need to systematically relax the errors by $\sim10\%$ of the measured fluxes.  This suggests that the quoted photometric uncertainties are systematically too constraining to allow a good fit given the parameter space our templates cover.  Relaxing the errors during fitting does not change our results; however, it is reflected in the reported $\chi_{\nu}^2$ of the fits.

The inclusion of an AGN template during SED fitting is preferred if the AGN template significantly improves the fit while ruling out the null hypothesis that this improvement is by random chance.  This is evaluated via the Fisher test through the F-ratio, which compares the reduced $\chi^2$ values and degrees of freedom of the galaxy only vs galaxy+AGN fits.  The interpretation of the F-ratio and its associated probability (F$_{\rm prob}$) is subject to the gaussian nature of the fit residuals as well as the rarity of the relevant population, which can lead to false positives.  To account for the latter effect, we calculate a corrected F$_{\rm prob}$ for all sources assuming a conservative 20$\%$ for the AGN fraction following the prescription in \citet{chu14}.  As this correction is a function of the number of degrees of freedom in the fit, it is minimal for most of our sources, which have $\sim17$ photometric bands.   A threshold of (corrected) F$_{\rm prob}\geq0.93$ is adopted to accept an AGN component in the fit.

\section{FSPS Spectra} \label{app:fsps}
To provide a representative base set of SFG models for the SED fitting of radio sources (Section~\ref{sec:sedfitting}), we generate 16 templates using the Flexible Stellar Population Synthesis stellar population code  \citep[FSPS v3.0; ][]{con09, con10}.  These templates have solar metallicity with an exponentially declining SFH ({\tt sfh=2}) based on a 2 Gyr old burst and e-folding time of 1 Gyr.  Dust content is treated self-consistently across the spectrum.  Dust attentuation is modeled according to \citet{cal00} with variable attenuation ($A_v=0-3$ in steps of 0.2).  Dust emission, dominated by PAH emission in the relevant wavelength range (0.03-30$\,\mu$m), is based on the \citet{dra07} models.  Dust emission parameters {\tt duste\_gamma=0.02}, {\tt duste\_umin=7}, and {\tt duste\_qpah=2} are chosen based on the stacked posteriors of energy-balanced fits to galaxies with {\it Herschel} photometry \citep{lej17}; the latter is chosen to be on the lower end of the distribution found in \citet{lej17} in order to provide a more reasonable 3.3$\,\mu$m feature, which is known to be overestimated in current modeling (B. Johnson, private communication).

%A handy "cheat sheat" that provides the necessary LaTeX to produce 17 
%different types of tables is available at %\url{http://journals.aas.org/authors/aastex/aasguide.html#table_cheat_sheet}.

%% The reference list follows the main body and any appendices.
%% Use LaTeX's thebibliography environment to mark up your reference list.
%% Note \begin{thebibliography} is followed by an empty set of
%% curly braces.  If you forget this, LaTeX will generate the error
%% "Perhaps a missing \item?".
%%
%% thebibliography produces citations in the text using \bibitem-\cite
%% cross-referencing. Each reference is preceded by a
%% \bibitem command that defines in curly braces the KEY that corresponds
%% to the KEY in the \cite commands (see the first section above).
%% Make sure that you provide a unique KEY for every \bibitem or else the
%% paper will not LaTeX. The square brackets should contain
%% the citation text that LaTeX will insert in
%% place of the \cite commands.

%% We have used macros to produce journal name abbreviations.
%% \aastex provides a number of these for the more frequently-cited journals.
%% See the Author Guide for a list of them.

%% Note that the style of the \bibitem labels (in []) is slightly
%% different from previous examples.  The natbib system solves a host
%% of citation expression problems, but it is necessary to clearly
%% delimit the year from the author name used in the citation.
%% See the natbib documentation for more details and options.

\bibliographystyle{aasjournal} % We choose the "plain" reference style
\bibliography{refs} % Entries are in the "refs.bib" file
%%\begin{thebibliography}{}

%%\end{thebibliography}

%% This command is needed to show the entire author+affilation list when
%% the collaboration and author truncation commands are used.  It has to
%% go at the end of the manuscript.
%\allauthors

%% Include this line if you are using the \added, \replaced, \deleted
%% commands to see a summary list of all changes at the end of the article.
%\listofchanges

\end{document}